\documentclass[11pt]{article}

\usepackage[preprint]{acl}

\usepackage{times}
\usepackage{latexsym}
\usepackage[T1]{fontenc}
\usepackage[utf8]{inputenc}
\usepackage{microtype}
\usepackage{inconsolata}
\usepackage{graphicx}

\usepackage{booktabs}        
\usepackage{multirow}
\usepackage{tabularx}
\usepackage{amsmath,amssymb} 
\usepackage[ruled,lined,linesnumbered,commentsnumbered]{algorithm2e}
\SetAlFnt{\small}
\usepackage{listings}        
\usepackage{CJKutf8}         
\usepackage{xcolor}
\usepackage{tikz}
\usetikzlibrary{positioning,arrows.meta}

\makeatletter
\def\input@path{{chapters/}{appendices/}{tables/}{figures/}{algorithms/}}
\makeatother
\graphicspath{{figures/}}

\newcommand{\ta}{Token-Aware}
\newcommand{\ws}{Whole-Sequence}

\title{Constrained Tagging for Span Labeling}

\author{Radek \v{S}tulc \\
  Faculty of Electrical Engineering \\
  Czech Technical University in Prague \\
  \texttt{stulcrad@alumni.cvut.cz} \And
  Jan Drchal \\
  Artificial Intelligence Center \\
  Czech Technical University in Prague \\
  \texttt{drchajan@fel.cvut.cz}}

\begin{document}
\maketitle

\begin{abstract}
Generative LLMs can label spans without task-specific training, but their output is not tied to positions in the input: a returned span may be absent from the input or occur several times. We present constrained tagging, a training-free method in which the model reproduces the input and wraps each span in labeled inline markup. A logits processor using a byte-level token trie and a finite-state machine admits only tokens that continue the input verbatim or form a valid tag, so the output reproduces the input exactly with labels from the given set, and spans may begin or end at any character boundary, even inside a token. Two-phase decoding lets reasoning models reason before the constraint engages. On six datasets and six open-weight models, the guarantee holds in all but 0.022\% of 255{,}312 constrained generations, each failure a reasoning model exhausting its token budget or never closing its reasoning, and the constraint adds no measurable cost. Unlike context-based prompting, which loses 8.5\% of its spans, constrained tagging locates every span it proposes on every dataset; it is as accurate on two datasets, and on the other four its errors lie in which spans the model chooses, not where they are.
\end{abstract}

\section{Introduction}
\label{sec:introduction}

Classifying text spans is a fundamental problem in NLP, with applications ranging from named entity recognition and extractive question answering to grammatical error correction, toxic language detection, and other information extraction tasks. Traditionally, this problem has been addressed with encoder-only models such as BERT \citep{devlin-etal-2019-bert}, which directly produce token-level predictions. These models need a large annotated dataset for every task, and they must be retrained whenever the label set changes. Generative large language models (LLMs), which have grown far larger and more capable and can now reason before they answer, offer a different paradigm: they can perform span labeling in a zero-shot setting through prompting alone, which makes them attractive when labeled data are scarce or the label set changes.

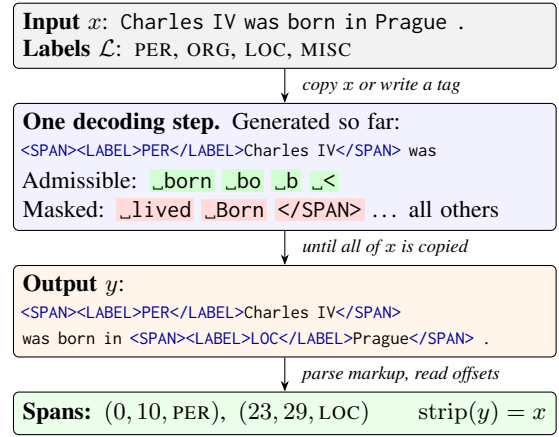
\begin{figure}[t]
\centering
\newcommand{\tg}[1]{\textcolor{blue!60!black}{#1}}
\newcommand{\ok}[1]{{\setlength{\fboxsep}{1pt}\colorbox{green!18}{\texttt{#1}}}}
\newcommand{\no}[1]{{\setlength{\fboxsep}{1pt}\colorbox{red!14}{\texttt{#1}}}}
\begin{tikzpicture}[
  font=\footnotesize, >={Stealth[length=4pt]}, node distance=4.5mm,
  box/.style={draw, rounded corners=2pt, align=left, text width=0.9\columnwidth, inner sep=3.5pt},
  arrowlabel/.style={font=\scriptsize\itshape, right=1mm, midway},
]
\node[box, fill=gray!10] (in) {%
  \textbf{Input} $x$:\enspace\texttt{Charles IV was born in Prague .}\\
  \textbf{Labels} $\mathcal{L}$:\enspace\textsc{per}, \textsc{org}, \textsc{loc}, \textsc{misc}};

\node[box, fill=blue!5, below=of in] (dec) {%
  \textbf{One decoding step.} Generated so far:\\
  {\scriptsize\texttt{\tg{<SPAN><LABEL>PER</LABEL>}Charles IV\tg{</SPAN>} was}}\\[2pt]
  Admissible:\enspace \ok{\textvisiblespace born}\ \ok{\textvisiblespace bo}\ \ok{\textvisiblespace b}\ \ok{\textvisiblespace<}\\[1pt]
  Masked:\enspace \no{\textvisiblespace lived}\ \no{\textvisiblespace Born}\ \no{</SPAN>}\ \ldots\ all others};

\node[box, fill=orange!8, below=of dec] (out) {%
  \textbf{Output} $y$:\\
  {\scriptsize\texttt{\tg{<SPAN><LABEL>PER</LABEL>}Charles IV\tg{</SPAN>}}}\\
  {\scriptsize\texttt{was born in \tg{<SPAN><LABEL>LOC</LABEL>}Prague\tg{</SPAN>} .}}};

\node[box, fill=green!8, below=of out] (spans) {%
  \textbf{Spans:}\enspace $(0, 10, \textsc{per})$,\enspace $(23, 29, \textsc{loc})$\hfill $\mathrm{strip}(y) = x$};

\draw[->] (in) -- node[arrowlabel]{copy $x$ or write a tag} (dec);
\draw[->] (dec) -- node[arrowlabel]{until all of $x$ is copied} (out);
\draw[->] (out) -- node[arrowlabel]{parse markup, read offsets} (spans);
\end{tikzpicture}
\caption{Constrained tagging. At every step, the processor admits only tokens that continue the input verbatim or belong to the tag markup (green) and masks all others (red), so the output reproduces $x$ exactly, and the offsets of each span follow from where its tag was generated.}
\label{fig:tagging}
\end{figure}

Applying generative LLMs to span labeling, however, introduces a challenge encoders do not face: \textbf{once the model identifies a span of interest, how do we locate it precisely within the original input text?} Generative LLMs operate auto-regressively, producing output tokens without any inherent correspondence to positions in the source text. This creates a fundamental localization problem, which is particularly acute when the same surface form appears several times: simply extracting ``Charles'' from the model output provides no information about \emph{which} occurrence was identified or \emph{where} in the text it is located (Fig.~\ref{fig:tagging}). Generative models may also fail to copy text faithfully, producing \emph{spans that cannot be aligned} to the source at all.

In this paper, we use a constrained decoding approach to span labeling. We classify spans with a \emph{tagging} mechanism: the model reproduces the input text and tags each span inline, wrapping it in markup that assigns it a label (Fig.~\ref{fig:tagging}). A logits processor, guided by a byte-level token trie \citep{fredkin-1960-trie} and a finite-state machine, restricts the generation at every step to tokens that either continue the input verbatim or form a valid tag with a label from the given label set. Localization is therefore enforced \emph{during} generation rather than recovered \emph{afterwards}: the position of every span is determined by construction at the moment it is generated, the output reproduces the input exactly, and a hallucinated span or an invalid label cannot be produced. Nothing is trained: adapting the method to a new task requires only a system prompt and a set of labels (\S\ref{sec:method}).

Our contributions are: (1) a training-free constrained decoding method for span labeling by tagging, which guarantees that the output reproduces the input verbatim with labels only from the given set, and places span boundaries at any character boundary, including inside a token (\S\ref{sec:method}); (2) two-phase decoding, which leaves the reasoning trace of a reasoning model unconstrained and constrains only its answer, so that reasoning models, including those that may skip their reasoning, keep the guarantee (\S\ref{sec:method-reasoning}); and (3) an evaluation on six datasets covering five span-labeling tasks, with six open-weight models from three families, against unconstrained tagging and context-based prompting, showing that the guarantee holds in all but 0.022\% of 255{,}312 constrained generations with no measurable overhead from the constraint, and that the remaining errors lie in which spans the model chooses rather than where they are (\S\ref{sec:experiments}).\footnote{Code and evaluation scripts: \url{https://github.com/stulcrad/constrained-tagging}}

\section{Background and Related Work}
\label{sec:related}

\paragraph{Encoder-based span classification.} Fine-tuned BERT \citep{devlin-etal-2019-bert} and its variants, such as RoBERTa \citep{liu-etal-2019-roberta}, achieved state-of-the-art results on many span classification tasks simply by placing a linear token classification head on top of the encoder's output representations. These models naturally provide position information, since the predicted label for token $i$ directly corresponds to position $i$ in the input. However, they require large labeled datasets for fine-tuning on each task and cannot adapt to new classes or domains without retraining, which motivates the exploration of generative LLMs.

\paragraph{Generative LLMs for span labeling.} \citet{ding-etal-2024-span} view span-oriented tasks as predicting triples of begin, end, and class, which puts one question at the center of every task: how to recover exact positions from the output of a generative model. GPT-NER \citep{wang-etal-2025-gpt} has the model reproduce the input with special tokens surrounding the entities of one type, while PromptNER \citep{ashok-lipton-2023-promptner}, CodeIE \citep{li-etal-2023-codeie}, InstructUIE \citep{wang-etal-2023-instructuie}, and UniversalNER \citep{zhou-etal-2024-universalner} return lists of entities without positions, as free text, code or the output of a fine-tuned model, so the spans must still be matched back to the input. \citet{dukic-snajder-2024-looking} instead make decoder-only models competitive with encoders by removing the causal mask in selected layers, which requires task-specific training. Surveys \citep{keraghel-etal-2024-recent,xu-etal-2024-large} consistently find that generative LLMs recognize what to extract but struggle with exact boundaries and alignment to the source: constraining the \emph{format} of the output does not constrain its \emph{content} to come verbatim from the input.

\paragraph{Strategies for span labeling.} Concurrently, \citet{semin-etal-2026-strategies} compare three ways a generative LLM can produce labeled spans: \emph{tagging} reproduces the input and wraps each span in markers, \emph{indexing} outputs numeric positions, which requires accurate counting, and \emph{matching} outputs the span text and aligns it to the source, which is ambiguous when a surface form repeats. On four tasks, three of which we use, they find tagging to be the most robust. Their constrained method, LogitMatch, restricts the span text of a matching output to substrings of the input; constraining tagging the same way ran into tokenization issues at the boundaries between text, tags, and span content, and was abandoned. Our processor handles all of these boundaries with one byte-level prefix search, and its copy cursor fixes the position of every span, including repeated ones that matching cannot tell apart.

\paragraph{Grammar-based constrained decoding.} Grammar-constrained decoding represents the valid outputs as a formal language and restricts each step to tokens that keep the output within it. Outlines \citep{willard-louf-2023-efficient} compiles regular expressions and context-free grammars into an index of valid next tokens, XGrammar \citep{dong-etal-2025-xgrammar} compiles a context-free grammar into a byte-level pushdown automaton, and llguidance \citep{guidance-2024-llguidance} computes each mask by walking a byte-level trie of the vocabulary against an Earley parser. A grammar fixed in advance enforces a schema, but it cannot require the output to reproduce the input: the copy language $\{ww\}$ is a textbook example of a language that is not context-free. \citet{geng-etal-2023-grammar} show that input-dependent grammars that reproduce the input with inline brackets let off-the-shelf LLMs parse without fine-tuning; they sketch similar grammars for chunking and semantic role labeling, but operate on words and leave tokenization ambiguity to tokenizer-specific workarounds, which our byte-level trie removes.

\paragraph{Trie-based and copy-constrained decoding.} A trie \citep{fredkin-1960-trie} supports prefix search, finding every stored string that is a prefix of a given string in time proportional to its length, which suits constrained generation, where each step must find the vocabulary tokens that keep the output within a set of valid strings. GENRE \citep{decao-etal-2021-autoregressive} decodes entity names from a trie and, for entity linking, constrains a fine-tuned model to copy the input token by token while inserting mention markup; \citet{zhang-etal-2023-seq2seq} also copy the input with inline coreference markers through a copy action and a constrained beam search, and iNERD \citep{deusser-etal-2023-informed} draws the entity text token by token from the input. All of them fine-tune the model and work at the granularity of tokens. We search a byte-level trie of the vocabulary, the structure of llguidance's toktrie, with the remaining input bytes, which lets an off-the-shelf model place span boundaries anywhere, including inside a token.

\section{Datasets}
\label{sec:datasets}

\begin{table*}[t]
\centering\footnotesize
\setlength{\tabcolsep}{4pt}
\begin{tabular}{@{}llcclrrrrr@{}}
\toprule
 & & & & & \multicolumn{2}{c}{median chars} & & & \\
\cmidrule(lr){6-7}
Dataset & Task & Lang. & $|\mathcal{L}|$ & Span geometry & span & input & spans/ex. & \% empty & $N$ \\
\midrule
CoNLL-2003   & NER           &  1 & 4 & token-aligned             &  8 &  41 &  1.6 & 20.2 & 1{,}152 / 3{,}453 \\
Toxic Spans  & toxicity      &  1 & 1 & word-aligned              &  6 & 142 &  0.6 & 48.6 &           1{,}000 \\
LegalQAEval  & extractive QA &  1 & 1 & clause--paragraph         & 30 & 577 &  0.5 & 50.0 &           1{,}206 \\
UniversalNER & NER           & 12 & 3 & token-aligned             &  7 & 104 &  1.9 &  0.0 & 3{,}702 / 7{,}523 \\
WMT24 ESA-MT & MT errors     &  8 & 2 & sub-word--clause          &  9 & 231 &  2.2 &  0.0 &               867 \\
MultiGEC     & GEC           &  1 & 3 & incl.\ \emph{zero-length} &  3 & 869 & 14.1 &  0.0 &               504 \\
\bottomrule
\end{tabular}
\caption{The six span-labeling datasets, and the axes along which they differ.}
\label{tab:datasets}
\end{table*}

Table~\ref{tab:datasets} summarizes the six datasets and the axes along which they differ. One method is applied to all of them, with only the system prompt changing to describe the task and the label set.

\paragraph{Named Entity Recognition (NER)} The goal is to localize the named entities and classify their type. We use two datasets. CoNLL-2003 \citep{tjong-kim-sang-de-meulder-2003-introduction} is the long-standing benchmark, built from Reuters newswire and annotated with four entity types: Person (\textsc{per}), Organization (\textsc{org}), Location (\textsc{loc}), and Miscellaneous (\textsc{misc}); its test split holds 3{,}453 sentences. UniversalNER \citep{mayhew-etal-2024-universal} extends the same task to 18 treebanks in 12 languages with three coarse types, \textsc{per}, \textsc{org}, and \textsc{loc}, for 7{,}523 examples. NER is the only task whose spans are guaranteed to fall on word boundaries, and the loader represents both datasets as BIO tags over whitespace tokens.

\paragraph{Toxic Span Detection (TOX)} Given a social-media post already judged toxic at the comment level, the goal is to find the character spans that carry the toxicity. We use Toxic Spans \citep{pavlopoulos-etal-2022-detection}, part of which formed SemEval-2021 Task~5 \citep{pavlopoulos-etal-2021-semeval}: posts from the Civil Comments corpus annotated by three to five crowdworkers, where a character is gold when a majority marked it. We evaluate on the 1{,}000 posts of its test split as distributed on the Hugging Face Hub. There is a single label, \textsc{toxic}. Almost half of the posts (48.6\%) are toxic in tone without any majority-agreed span, so their gold is the empty set.

\paragraph{Extractive Question Answering (QA)} The goal is to locate the span of a passage that answers a question, so the admissible spans are set by the question rather than by a fixed schema. We use LegalQAEval \citep{butler-2025-legalqaeval}, 1{,}206 test items whose legal passages were filtered out of SQuAD~2.0, MS MARCO, HotpotQA and Natural Questions. As in TOX, there is a single label, \textsc{answer}. Exactly half of the items are unanswerable from their passage and carry empty gold.

\paragraph{Detecting Errors in Machine Translation (ESA-MT)} The goal is to find error spans in a translation given its source. The gold is the human error-span annotation collected at WMT24 \citep{kocmi-etal-2024-findings} under the ESA protocol of \citet{kocmi-etal-2024-error}; we read it from the WMT24 sample released by \citet{kasner-etal-2026-llms}, restricted to the examples that contain at least one error. That leaves 867 examples over 24 source--target--domain files, covering translation out of English into eight languages across the news, social, and literary domains. Following the annotation scheme, spans carry a severity, either \textsc{minor} or \textsc{major}. The source is shown to the model as context and is never copied or tagged.

\paragraph{Grammatical Error Correction (GEC)} The goal is to identify the spans that require editing to make the text correct. We use the English Write \& Improve portion \citep{nicholls-etal-2024-write} of MultiGEC \citep{masciolini-etal-2025-multigec}, 504 learner essays whose edits are typed by ERRANT \citep{bryant-etal-2017-automatic} and grouped into three high-level categories: replacement (\textsc{r}), missing (\textsc{m}), and unnecessary (\textsc{u}). In GEC, the \textsc{m} label marks a position where the text is missing, so the gold spans for \textsc{m} are of zero length. The other two labels, \textsc{r} and \textsc{u}, always have non-zero length.

\paragraph{Subsampling.} Two datasets are subsampled to fit our compute budget: CoNLL-2003 to 1{,}152 of its 3{,}453 test sentences, and UniversalNER to 250 examples per treebank (3{,}702 of 7{,}523; four of the 18 treebanks are smaller than 250 and are used whole). We draw the subsample uniformly at random and re-draw it per seed, so the three seeds vary the evaluation set as well as the decoding; within a seed, every system compared on that dataset sees the same examples. We fully evaluate the remaining four datasets.

\paragraph{Normalization.} Inputs are Unicode NFC-normalized in the loader, with gold offsets remapped, and are otherwise left exactly as shipped, since the model must reproduce them.

\section{Methodology}
\label{sec:method}

\subsection{Problem Statement}
\label{sec:problem}

Given an input text $x$ and a label set $\mathcal{L}$, span labeling asks for the set of labeled spans $S = \{(s_k, e_k, \ell_k)\}$, where $\ell_k \in \mathcal{L}$ and $0 \le s_k \le e_k \le |x|$ are offsets between the characters of $x = x_1 \dots x_{|x|}$, so the span covers $x_{s_k+1} \dots x_{e_k}$. The set may be empty, and a span may have zero length ($s_k = e_k$) when it marks a position in the text rather than any text at all.

We solve this task by \emph{constrained tagging}: the model is prompted to return $x$ with every span it identifies written as
\begin{center}
\texttt{<SPAN><LABEL>}$\ell$\texttt{</LABEL>}\textit{span text}\texttt{</SPAN>}
\end{center}
with $\ell \in \mathcal{L}$ (the prompts are in Appendix~\ref{app:prompts}). A valid output $y$ must satisfy $\mathrm{strip}(y) = x$, where $\mathrm{strip}$ removes the markup, and every label in $y$ must belong to $\mathcal{L}$. The processor advances a cursor through $x$ as the model copies it, so the offsets of every span follow from the position of the cursor when its tag opens, and repeated surface forms need no alignment step. After generation, the markup is parsed into labeled spans with character offsets; without the constraint, the share of outputs that fail $\mathrm{strip}(y) = x$ is the wrong-text rate (\S\ref{sec:experiments}).

Both conditions are enforced at decoding time by a logits processor that masks every token that would violate them. A byte-level token trie supplies the tokens that continue $x$ from the current position (\S\ref{sec:method-trie}), a finite-state machine decides where markup may appear and which labels it may carry (\S\ref{sec:method-processor}), and a two-phase extension lets reasoning models reason before the constraint engages (\S\ref{sec:method-reasoning}). The pseudocode for building and searching the trie is given in Appendix~\ref{app:algorithms}.

\subsection{Byte-Level Token Trie}
\label{sec:method-trie}

Span boundaries belong to the text, not its tokenization, so we allow a span to begin or end at any character boundary of $x$. This matters in practice: in ESA-MT, 22\% of error spans begin or end inside a word, typically at an incorrect inflectional ending, and nothing guarantees the tokenizer splits the word there. Copying is therefore not tied to the input's tokenization: a trie over the vocabulary returns every token that is a prefix of the remaining input, including tokens that end inside an input token. The trie is keyed on bytes, because in byte-level vocabularies a token can hold only part of a multi-byte character and has no character string of its own. Markup may still open or close only at a character boundary (guard~(i), \S\ref{sec:method-processor}); the bytes inside a character are only copied through.

The trie is built once per tokenizer and reused across decoding steps and examples. Each vocabulary token is inserted along the path of its bytes, and the node where the path ends stores its id. The bytes are read directly from the token's vocabulary entry, for both byte-level BPE \citep{sennrich-etal-2016-neural} and SentencePiece \citep{kudo-richardson-2018-sentencepiece} with byte fallback (Alg.~\ref{alg:trie-build}). They are never obtained by decoding the token to text and encoding it again, because a token that holds only part of a multi-byte character does not decode to valid text on its own.

The copy source is built the same way: $x$ is tokenized, and the bytes of its tokens are concatenated. The processor checks that the result equals the UTF-8 encoding of $x$, since a tokenizer that normalizes its input would otherwise make the copy target differ from the text being labeled; this is why inputs are NFC-normalized in the loader (\S\ref{sec:datasets}). The positions in $x$ are byte offsets into this encoding. The processor precomputes which of them are character boundaries and allows markup only at those. A prefix search walks the remaining input bytes from the root and collects the ids stored at every node it passes, stopping at the first byte with no child (Alg.~\ref{alg:prefix}). Its cost is bounded by the length of the longest vocabulary token and does not depend on the length of $x$.

\subsection{Constrained Logits Processor}
\label{sec:method-processor}

\begin{figure}[t]
\centering
\includegraphics[width=\columnwidth]{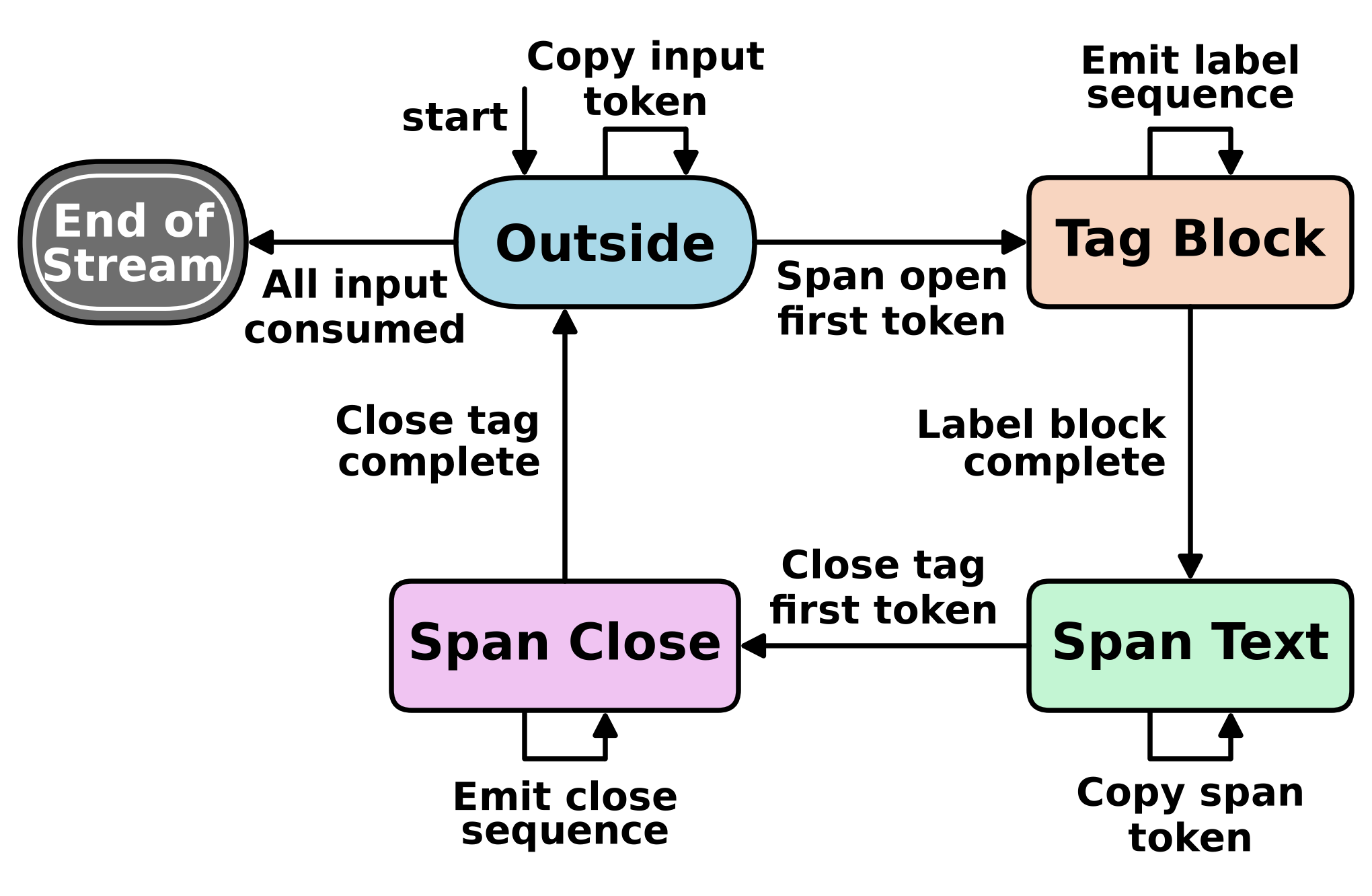}
\caption{The finite-state machine of the processor. Every edge is a set of admissible tokens; the machine accepts only after all of the input has been copied.}
\label{fig:fsm}
\end{figure}

The processor is a finite-state machine with four states (Fig.~\ref{fig:fsm}). At every step, it advances the machine by the just-generated token, computes the next admissible tokens, and sets the logits of all other tokens to $-\infty$, so that sampling renormalizes over the admissible set. In the copying states, \textsc{outside} and \textsc{span}, the admissible set is the trie's tokens for the remaining input, plus the first token of an opening block, or of \texttt{</SPAN>} once the span holds text; end-of-sequence becomes admissible only after all of $x$ has been copied. In \textsc{tag}, a label remains admissible only while its pre-tokenized opening block agrees with every token emitted since the tag opened, so the label that completes is always one of $\mathcal{L}$.

Tags are written the way the model's own tokenizer writes them. BPE tokenizers merge the space before a word into that word's first token, so after a space, a model emits the space and the \texttt{<} of a tag as one token, not two. Each opening block is therefore pre-tokenized with and without a leading space, and when the next input byte is a space, the spaced variant is offered: the tag absorbs the space, the cursor moves past it, and the space is kept in the stripped output.

Three guards close failure modes that each produced plausible but wrong output rather than an error. (i)~Markup may open or close only at a codepoint boundary. Without this rule, a span can close between the bytes of a single character, leaving two invalid fragments. (ii)~The tags are plain text rather than special tokens, so a literal \texttt{<} in the input is the same token as the start of a tag. Where the input continues with \texttt{<}, no tag may open, and the bracket is copied rather than read as markup; this keeps a span from starting exactly at a literal bracket, which 2 of the roughly 19{,}200 gold spans in our data do. (iii)~Zero-length spans are opt-in per label. A span whose label is in $\mathcal{E}_{\text{allow}}$ may close before any text is copied, and one whose label is in $\mathcal{E}_{\text{force}}$ must; in MultiGEC, \textsc{m} is forced and every other label needs at least one character. So that empty spans cannot repeat without end, no two may close at the same position, and a forced span opened where an empty span has just closed copies text instead.

The copy cursor tracks the current input token and the byte offset inside it, and the trie is searched only with the rest of that token. A copied token can therefore end inside an input token, so every boundary stays reachable, but it never crosses into the next one: the copy keeps to the segmentation of the input that the model saw in its prompt.

Every answer token is admissible, so the answer is always a well-formed tagging of a prefix of $x$ whose labels belong to $\mathcal{L}$, and end-of-sequence becomes admissible only once all of $x$ has been copied. The output therefore reproduces $x$ exactly, unless the generation is cut off first by the token budget or by the dead-end fallback (Appendix~\ref{app:processor}).

\subsection{Two-Phase Reasoning-Aware Decoding}
\label{sec:method-reasoning}

A processor that masks every step also masks the reasoning trace. As in \citet{nguyen-etal-2026-thinking}, who apply structured decoding only after a trigger token, and CRANE \citep{banerjee-etal-2025-crane}, which switches a grammar constraint on and off at delimiters, we therefore decode in two phases: the model first reasons without constraint, and the constraint governs the answer from the family-specific marker that closes the reasoning block. One case needs attention. If the chat template leaves the reasoning opener to the model, the model may answer directly, and a processor waiting for the closing marker would leave the whole answer unconstrained; for such models, the first step admits the opener together with the constrained answer tokens.

The guarantee is therefore conditional: $x$ is reproduced exactly whenever the model closes its reasoning. A model that exhausts the token budget or ends its turn while still reasoning leaves an empty answer, which is credited with no spans and counts toward the wrong-text rate; Table~\ref{tab:failures} in the appendix lists every such case.

\subsection{Complexity}
\label{sec:method-complexity}

Building the trie is linear in the total byte length of the vocabulary and happens once per tokenizer. At each step, the prefix search is bounded by the length of the longest token, and the remaining checks take constant time or time linear in $|\mathcal{L}|$, so masking the vocabulary, $\mathcal{O}(|\mathcal{V}|)$, dominates. Over $G$ generated tokens, constrained decoding therefore costs $\mathcal{O}(G \cdot |\mathcal{V}|)$, the same order as unconstrained decoding, and it does not depend on the length of the input.

\section{Experiments and Results}
\label{sec:experiments}

\paragraph{Setup.} We evaluate six open-weight models from three families: gemma-4-E2B-it and gemma-4-31B-it \citep{gemmateam-2026-gemma4}, Qwen3-8B \citep{qwenteam-2025-qwen3} and Qwen3.8-27B \citep{qwenteam-2026-qwen38}, and gpt-oss-20b and gpt-oss-120b \citep{openai-2025-gptoss}, each with reasoning off where it can be switched off and four of them also with more reasoning. We sample with each vendor's recommended parameters and a budget of 18{,}000 new tokens, at batch size~1 and, on the two NER datasets, also at batch size~64, and report the mean over three seeds (Appendix~\ref{app:setup}). Batch size~64 simulates a long document with many spans to find: unlike the usual sense of the term, the batch is not decoded in parallel, but its 64 sentences are concatenated into one input, with over a hundred entities on average, which the model copies and tags in a single generation.

\paragraph{Baselines.} We compare constrained tagging (\S\ref{sec:method}) against three baselines. \emph{Unconstrained tagging} uses the same prompt and markup with the processor switched off; an output whose stripped text differs from the input cannot be located and is credited with no spans. \emph{Context-based prompting} asks the model for a JSON list of spans, each with its label and a short context copied from the input, and locates each span by exact matching of its context. \emph{XGrammar} \citep{dong-etal-2025-xgrammar} also enforces the JSON format and the set of labels using a grammar. Due to limited computational resources, context-based prompting was run for one model per family and XGrammar only with reasoning off (Appendix~\ref{app:setup}).

\paragraph{Metrics.} Our headline metric is the character-overlap F1 of \citet{semin-etal-2026-strategies}, computed with their implementation: \emph{hard} F1 requires the predicted label to match, and \emph{soft} F1 credits a span in the right place whatever its label. Overlap counts are pooled over all examples, so an example with empty gold and an empty prediction carries no weight, while spans predicted where the gold is empty still lower precision. This keeps the metric informative on Toxic Spans and LegalQAEval, where half of the gold is empty, and the per-example character F1 of \citet{pavlopoulos-etal-2021-semeval} reaches 50\% for a model that never returns a usable span. UniversalNER and ESA-MT are macro-averaged over their subsets. The \emph{wrong-text rate} is the share of generations whose stripped output is not the input.

\begin{table*}[t]
\centering\footnotesize
\setlength{\tabcolsep}{5pt}
\begin{tabular}{@{}l cccccc@{}}
\toprule
& CoNLL & Toxic Spans & LegalQAEval & UNER$^{\dagger}$ & ESA-MT$^{\dagger}$ & MultiGEC \\
\midrule
\multicolumn{7}{@{}l}{\emph{Constrained tagging (ours)}} \\
gemma-4-E2B-it (off)    & 56.4 (76.7) & 13.2 & 11.2 & 58.6 (66.3) & 4.6 (9.8)   & 8.2 (12.8) \\
gemma-4-31B-it (off)    & \textit{84.7} (94.6) & 25.3 & \textit{55.5} & \textit{88.7} (92.2) & \textbf{14.0} (34.0) & \textit{50.7} (54.0) \\
Qwen3.8-27B (off)       & 72.8 (81.5) & 17.4 & 41.8 & 59.8 (63.2) & 9.2 (22.8)  & 24.5 (31.1) \\
Qwen3.8-27B (low)       & 80.4 (88.5) & 20.0 & 51.8 & 54.4 (56.7) & 9.5 (25.9)  & 49.9 (52.7) \\
gpt-oss-20b (low)       & 69.9 (81.9) & 20.5 & 42.9 & 75.7 (80.4) & 12.9 (25.7) & 24.5 (28.1) \\
gpt-oss-120b (low)      & 78.6 (89.6) & 20.6 & 47.9 & 79.5 (83.1) & 11.8 (28.2) & 31.8 (36.4) \\
\midrule
\multicolumn{7}{@{}l}{\emph{Unconstrained tagging (same prompt, constraint off)}} \\
gemma-4-E2B-it (off)    & 11.3 (14.1) & 28.7 & 2.1  & 22.2 (24.9) & 2.0 (5.0)   & 0.7 (0.8) \\
gemma-4-31B-it (off)    & 72.4 (81.7) & 24.9 & \textbf{56.0} & 84.5 (87.9) & \textit{13.9} (34.1) & 36.3 (38.5) \\
Qwen3.8-27B (off)       & 69.9 (78.6) & 23.3 & 46.4 & 60.7 (64.2) & 9.1 (22.3)  & 31.1 (32.7) \\
Qwen3.8-27B (low)       & 83.0 (90.9) & 21.2 & 52.9 & 70.6 (73.2) & 10.0 (25.1) & 46.2 (48.3) \\
gpt-oss-20b (low)       & 59.8 (72.0) & 23.7 & 40.0 & 70.0 (73.4) & 10.2 (20.7) & 14.2 (15.3) \\
gpt-oss-120b (low)      & 72.9 (83.9) & 21.9 & 46.8 & 77.9 (81.4) & 10.4 (24.3) & 17.6 (18.5) \\
\midrule
\multicolumn{7}{@{}l}{\emph{Context-based prompting}} \\
gemma-4-E2B-it (off)    & 71.5 (85.9) & \textbf{32.5} & 38.6 & 78.0 (82.9) & 6.5 (17.2)  & 27.0 (29.0) \\
gemma-4-E2B-it (on)     & 73.1 (85.9) & 29.5 & 40.3 & 83.9 (88.0) & 9.0 (20.1)  & 33.9 (35.8) \\
Qwen3.8-27B (off)       & 84.1 (92.5) & 20.4 & 17.4 & 86.9 (90.3) & 8.8 (20.7)  & 40.1 (43.0) \\
Qwen3.8-27B (low)       & \textbf{87.4} (95.7) & 28.4 & 33.5 & \textbf{89.4} (92.5) & 10.1 (26.4) & \textbf{51.6} (53.8)  \\
gpt-oss-20b (low)       & 73.3 (84.4) & \textit{30.0} & 32.1 & 80.9 (85.0) & 11.1 (24.5) & 32.7 (34.9) \\
\bottomrule
\end{tabular}
\caption{Hard F1 (soft F1 in parentheses) on the six datasets, mean over three seeds, batch size~1. $^{\dagger}$Macro average over the 18 treebanks and the 24 files. The parenthesized arm is the reasoning setting. The best hard F1 per dataset is in bold, the second best in italics.}
\label{tab:main}
\end{table*}

\subsection{Main Results}
\label{subsec:main-results}

Table~\ref{tab:main} gives the results on the six datasets. Constrained tagging is ahead of unconstrained tagging, with the same prompts, on five of the six datasets, and behind only on Toxic Spans (Table~\ref{tab:full} in the appendix). The gain is largest where a model fails to copy its input: gemma-4-E2B-it rises from 11.3 to 56.4 hard F1 on CoNLL-2003, and gemma-4-31B-it from 36.3 to 50.7 on MultiGEC.

Against context-based prompting, the strongest baseline, constrained tagging is on par on average on LegalQAEval (+1.0) and ESA-MT (+0.1) and behind on the other four datasets. The gap is largest for the smallest model, gemma-4-E2B-it; without it, constrained tagging is ahead on every LegalQAEval arm, by up to 24.4 points for Qwen3.8-27B, and on three of four ESA-MT arms, and at medium effort gpt-oss-20b is ahead on four of the six datasets, including CoNLL-2003 and UniversalNER. Where constrained tagging stays behind, \S\ref{subsec:analysis} traces the gap to which spans the model chooses, not to where they are. XGrammar changes the context-based results by at most 1.8 points in either direction: constraining the form of the output does not change what the model extracts. Standard deviations over seeds are at most 2.7 points for any constrained arm (Table~\ref{tab:full} in the appendix).

\begin{table}[t]
\centering\footnotesize
\setlength{\tabcolsep}{4pt}
\begin{tabular}{@{}lccc@{}}
\toprule
& UNER & MultiGEC & ESA-MT \\
\midrule
\multicolumn{4}{@{}l}{\emph{gemma-4-31B-it}} \\
Published tagging     & \textit{88.9} & \textbf{52.9} & \textbf{14.2} \\
Published LogitMatch  & \textbf{89.2} & 41.7 & 13.1 \\
Ours, unconstrained   & 84.7 & 36.3 & 13.9 \\
Ours, constrained     & 88.8 & \textit{50.7} & \textit{14.0} \\
\midrule
\multicolumn{4}{@{}l}{\emph{Qwen3-8B}} \\
Published tagging     & \textbf{73.3} & \textbf{27.2} & 10.2 \\
Published LogitMatch  & \textbf{73.3} & \textit{17.4} & 11.2 \\
Ours, unconstrained   & 62.2 & 13.8 & \textit{13.3} \\
Ours, constrained     & \textit{70.2} & 15.7 & \textbf{13.6} \\
\bottomrule
\end{tabular}
\caption{Hard F1 against the published tagging and LogitMatch results of \citet{semin-etal-2026-strategies}, reasoning off, recomputed from their released runs; ours over three seeds, theirs over five, UNER on the full test set. Per model, the best is in bold, the second best in italics.}
\label{tab:published}
\end{table}

\paragraph{Published results.} Table~\ref{tab:published} compares against \citet{semin-etal-2026-strategies} on the two models evaluated by both works. With gemma-4-31B-it, constrained tagging matches their tagging results within 0.2 points on UniversalNER and ESA-MT, although our scoring of outputs that do not reproduce the input is stricter; on MultiGEC it is 2.2 points below their tagging but 9.0 points ahead of their constrained matching method, LogitMatch. With Qwen3-8B, it is ahead of both on ESA-MT and behind on the other two datasets.

\begin{table}[t]
\centering\footnotesize
\setlength{\tabcolsep}{4pt}
\begin{tabular}{@{}lcccc@{}}
\toprule
\% & Context & XGr. & Unconstr. & Ours \\
\midrule
Span not in input  & 8.53 & 9.22 & ---   & 0 \\
Label not in set   & 0.01 & 0.00 & ---   & 0 \\
Wrong text         & ---  & ---  & 34.61 & 0.01 \\
Unusable output    & 0.54 & 0.01 & 34.61 & 0.01 \\
\midrule
No answer          & 0.02 & ---  & 0.01  & 0.01 \\
\bottomrule
\end{tabular}
\caption{Failure modes, averaged over the six datasets, on the arms run for every method. XGr.: XGrammar; Unconstr.: unconstrained tagging. Span and label rows are in \% of proposed spans, the others in \% of generations; \emph{No answer} means the reasoning never closed.}
\label{tab:errors}
\end{table}

\subsection{The Verbatim Guarantee}
\label{subsec:guarantee}

Table~\ref{tab:errors} lists how each method fails. Under the constraint, a span that does not occur in the input and a label outside the label set cannot be produced, and a generation can fail only by not finishing its answer, either because it reaches the budget of 18{,}000 new tokens or because the model ends its turn while still reasoning. Of the 255{,}312 constrained generations, 56 (0.022\%) failed: in the ten at batch size~1, the reasoning never closed, and of the 46 at batch size~64, 28 ran out of budget during the answer (Table~\ref{tab:failures} in the appendix). Without the constraint, the same prompts produce wrong text in 34.6\% of generations, which then lose all of their spans, and context-based prompting returns 8.5\% of its spans with a context that cannot be found in the input, 22.4\% on LegalQAEval.

\subsection{Where Accuracy Is Lost}
\label{subsec:analysis}

The guarantee fixes span location, not span selection or classification, and these are where the remaining errors lie (Tables~\ref{tab:spans} and~\ref{tab:confusion} in the appendix). Reproducing the whole input takes the model past every candidate, and at each one it must decide whether to open a span. It errs toward opening one: of the three methods, constrained tagging misses the fewest gold spans on every dataset, but it proposes more spans than context-based prompting on five of the six and longer ones on all six, 138 characters on LegalQAEval against 63 in the gold. The recall is paid for in precision, since fewer of its spans are exact on five datasets and more are spurious or misdelimited: on CoNLL-2003, 12.8\% of its spans have the right label but wrong boundaries, compared to 3.7\% for context-based prompting. The constraint thus removes errors in reproducing the input but leaves errors in judging which text is a span, and no constraint on the output can remove those. The same holds for labels: 31.7\% of the correctly delimited \textsc{org} spans of CoNLL-2003 are labeled \textsc{loc}, and on ESA-MT 71.8\% of the \textsc{minor} errors are labeled \textsc{major}, which is why hard F1 there is a fraction of soft F1 (14.0 against 34.0 for gemma-4-31B-it).

\subsection{Effect of Reasoning}
\label{subsec:reasoning}

More reasoning improves the hard F1 averaged over the six datasets for every model, by 2.6 to 6.7 points, but not on every dataset (Table~\ref{tab:reasoning} in the appendix): the largest gain of each model is on MultiGEC for Qwen3.8-27B (+25.4), on CoNLL-2003 for gemma-4-E2B-it (+11.6) and on UniversalNER for gpt-oss-20b and gpt-oss-120b (+10.5 and +9.8), while gpt-oss loses on Toxic Spans ($-$3.5 and $-$1.7) and Qwen3.8-27B on UniversalNER ($-$5.4). The cost is high: gpt-oss-20b spends 1{,}290 reasoning tokens per generation at medium effort against 207 at low (Fig.~\ref{fig:tokens} in the appendix). Under the two-phase processor, reasoning fails to close in at most 0.13\% of an arm's generations at batch size~1, but in up to 5.5\% at batch size~64.

\subsection{Cost and Batch Size}
\label{subsec:cost}

The constraint itself costs nothing measurable: constrained and unconstrained tagging, run in the same jobs, generate 19.5 and 19.4 tokens per second averaged over all models and datasets. Tagging pays instead for reproducing the whole input. Without reasoning, it is therefore slower than context-based prompting where a long input holds few spans, as in Toxic Spans and LegalQAEval with 0.6 and 0.5 spans per example (Table~\ref{tab:datasets}), and faster on MultiGEC, where 14 spans per essay make the list of spans the longer output (Appendix~\ref{app:runtime}). Batch size~64 strains both methods (Tables~\ref{tab:batch} and~\ref{tab:batch-recall} in the appendix): most models propose fewer spans, and recall falls. With gemma-4-E2B-it and gpt-oss-20b, constrained tagging keeps more of the gold spans on UniversalNER than context-based prompting, which recovers as little as 17.6\% of them, and for gpt-oss-20b this makes it the more accurate method at batch size~64 (43.2 against 23.5 hard F1 at low effort, 53.7 against 44.5 at medium). For gemma-4-E2B-it the extra spans are mostly wrong, and on CoNLL-2003 its hard F1 drops from 56.4 to 8.5, against 50.5 for context-based prompting.

\section{Conclusion}
\label{sec:conclusion}

We presented constrained tagging, a training-free decoding method that guarantees by construction that every span a generative LLM returns is a verbatim part of the input with a label from the given set. The guarantee held in all but 0.022\% of 255{,}312 generations and pays off where text-based localization breaks down: with the same prompts, it raises hard F1 over unconstrained tagging by up to 45 points, it never loses a span it proposes, whereas context-based prompting loses up to 22\% on long passages, and it matches context-based prompting on legal question answering and translation error detection. Because tagging marks spans in place, it suits densely annotated text: on MultiGEC, with 14 spans per essay, its output is shorter than a list of spans, and with gemma-4-31B-it it is 9.0 points ahead of a published constrained matching method. Of the three methods, it also misses the fewest gold spans on every dataset, which is what matters when a model proposes spans for a human annotator to accept or reject. Its remaining errors concern which spans the model chooses, which better prompts or light fine-tuning may reduce without touching the guarantee.

\section*{Limitations}

\paragraph{The guarantee is conditional.} Constrained tagging reproduces the input exactly only if the model closes its reasoning and finishes its answer within the token budget. Reasoning effort and batch size therefore have to be chosen with the budget in mind, and a failed generation yields no spans at all.

\paragraph{The guarantee does not choose better spans.} Constrained tagging removes spans that cannot be located, but on four of the six datasets context-based prompting with the same models is more accurate (\S\ref{subsec:main-results}). The verbatim guarantee concerns where the predicted spans are, not whether they are the right spans with the right labels; the errors that remain are analyzed in \S\ref{subsec:analysis}.

\paragraph{Logit access is required.} The processor masks the logits at every decoding step, so it needs a model whose decoding loop can be modified. It cannot be applied through a chat API that returns only text, whereas the context-based baseline can.

\paragraph{The cost grows with the input.} Tagging regenerates the whole input, so the length of its answer grows with the input rather than with the number of spans. Without reasoning, it therefore generates 3.6 times as many tokens as context-based prompting on LegalQAEval, whose long passages hold few answer spans. With reasoning, both methods also generate a reasoning trace, which narrows the ratio to 1.3. The constraint itself adds no measurable decoding cost (\S\ref{subsec:cost}).

\paragraph{Sampling and statistical power.} CoNLL-2003 and UniversalNER are subsampled to 1{,}152 sentences and 250 examples per treebank, drawn uniformly at random and re-drawn per seed, so their standard deviations include subsampling variance. All results are means over three seeds. This is enough to separate the large differences we report, but no significance tests were run, and differences of a point or two between two arms should not be read as established.

\paragraph{Not every configuration was run.} Due to computation, the prompting baselines were run for one model per family and XGrammar only with reasoning off; gemma-4-31B-it and Qwen3-8B were run with reasoning off only; and gpt-oss cannot switch reasoning off at all, so its reasoning comparison is between two efforts. The \ws{} processor was evaluated only in earlier experiments with smaller models, and no fine-tuned model was trained; that light fine-tuning could close the gap to task-specific encoders is a hypothesis we did not test.

\paragraph{Inherited data conventions.} Gold spans, label sets, and averaging follow the released benchmarks, including the subset-level macro averages of UniversalNER and ESA-MT, the empty gold of half of Toxic Spans and LegalQAEval, and 19 zero-length MultiGEC spans that lie one character past the end of their text and cannot be reached by any method. We kept them so that our results stay comparable with published ones. Inputs are NFC-normalized, so the verbatim guarantee holds relative to that normal form.

\paragraph{Prompts and coverage.} Each dataset uses one English system prompt, the same for every model, written once and not tuned per model, and the results may change with other prompts. The evaluation covers five tasks and 18 languages in five scripts, which shows breadth, not universality.

\paragraph{Contamination.} All six datasets are public, and the models were pre-trained on web text after their release, so we cannot exclude that test data was seen during pre-training. This affects all methods compared here alike, but it may inflate the absolute scores.

\section*{Ethics Statement}

\paragraph{Data and licences.} All six datasets are public and were used only for research evaluation; no model was trained on them, and no new annotation was collected. CoNLL-2003 annotates Reuters news stories, which are copyrighted and distributed for research under the Reuters Corpus agreement; we used the copy on the Hugging Face Hub. Toxic Spans is released under CC0. LegalQAEval is released under CC BY-SA 4.0 and inherits the terms of its sources, including the non-commercial terms of MS MARCO. The UniversalNER treebanks are licensed individually: twelve under CC BY-SA 4.0, the five PUD treebanks under CC BY-SA 3.0, and Tagalog-Ugnayan under CC BY-NC-SA 4.0. The WMT24 data are released for research use, and we read the error-span sample from the repository of \citet{kasner-etal-2026-llms}, released under the MIT license. The English Write \& Improve 2024 corpus behind MultiGEC is provided by Cambridge University Press \& Assessment under a license for non-commercial research and education that forbids redistribution; we acknowledge the licensor here, and our released code does not include the corpus: users obtain it from the licensor, and our scripts convert it. The character-overlap metric of \citet{semin-etal-2026-strategies} is imported from their public repository at a pinned commit and not redistributed, because that repository does not carry a license. All six models are released under the Apache 2.0 license, subject to their providers' usage policies, and our use is consistent with these terms.

\paragraph{Offensive and personal content.} Toxic Spans consists, by construction, of toxic user comments, and its system prompts, released with our code, quote two short examples of such language, one of them an insult aimed at a public figure. We did not process personal data beyond what the published datasets contain.

\paragraph{Intended use and risks.} We propose constrained tagging as an assistant to human annotators, not as an autonomous labeler. It guarantees that every predicted span is a verbatim part of the input with a valid label; it is not a guarantee that the spans or labels are correct. A user who reads the verbatim guarantee as a correctness guarantee would trust wrong labels and boundaries, which remain common (\S\ref{subsec:analysis}), so outputs used for consequential decisions, for example, in legal question answering, should be reviewed by a person.

\paragraph{Compute.} The experiments used about 4{,}300 GPU-hours, including pilot runs and failed jobs (Appendix~\ref{app:setup}).

\section*{Acknowledgments}
The access to the computational infrastructure of the OP VVV funded project CZ.02.1.01/0.0/0.0/16\_019/0000765 ``Research Center for Informatics'' is gratefully acknowledged. We used an AI assistant (Claude) to write analysis and plotting code, to adapt the authors' own earlier write-up into paper sections, to draft some passages that the authors then revised, for language editing, and to verify citations; the authors checked all content and take full responsibility for it.

\bibliography{custom}

\appendix
\section{Experimental Setup}
\label{app:setup}

\begin{table}[t]
\centering\footnotesize
\setlength{\tabcolsep}{4pt}
\begin{tabular}{@{}lll@{}}
\toprule
Model & Parameters & Reasoning arms \\
\midrule
gemma-4-E2B-it & 4.6B (2.3B eff.) & off, on \\
gemma-4-31B-it & 30.7B & off \\
Qwen3-8B       & 8.2B  & off \\
Qwen3.8-27B    & 27B   & off, low \\
gpt-oss-20b    & 20.9B (3.6B act.) & low, medium \\
gpt-oss-120b   & 116.8B (5.1B act.) & low, medium \\
\bottomrule
\end{tabular}
\caption{Evaluated models, with the parameters of the language model only, without vision or audio encoders. The Gemma and Qwen models are dense; gpt-oss is a mixture of experts. eff.: effective parameters (per-layer embeddings excluded); act.: parameters active per token. gpt-oss cannot switch reasoning off, so its arms are reasoning efforts.}
\label{tab:models}
\end{table}

\paragraph{Models.} Table~\ref{tab:models} gives the sizes of the six models of \S\ref{sec:experiments} and their reasoning settings. gemma-4-E2B-it is additionally run with reasoning on, and Qwen3.8-27B with low reasoning effort; gpt-oss cannot switch reasoning off and is run at low and medium effort. gemma-4-31B-it and Qwen3-8B are run with reasoning off only, the setting of the published numbers we compare against.

\paragraph{Decoding.} All runs sample with the parameters each vendor recommends: temperature 1.0, top-$p$ 0.95 and top-$k$ 64 for Gemma; temperature 0.6, top-$p$ 0.95 and top-$k$ 20 for Qwen with reasoning, and 0.7, 0.8 and 20 without it; temperature 1.0 and no truncation for gpt-oss. No repetition penalty is applied. The budget is 18{,}000 new tokens per generation. Every dataset is evaluated at batch size~1, and the two NER datasets also at batch size~64, where the 64 sentences of a batch are tagged in one generation. Constrained and unconstrained tagging use the same system prompt and differ only in whether the processor is attached; both decoding modes of an arm run in the same job on the same hardware.

\paragraph{Seeds.} Every configuration is run with seeds 42, 43, and 44, and every reported number is the mean over the three, with the standard deviation over seeds as the error. The seed fixes both the decoding and, on CoNLL-2003 and UniversalNER, the evaluation subsample (\S\ref{sec:datasets}).

\paragraph{Baselines.} The context-based prompt asks for a JSON list of spans, each with its label and a short surrounding context copied from the input (four to eight words; three to eight on Toxic Spans), using the same task description as the tagging prompt; a deterministic algorithm locates each span by exact matching of its context in the input. The XGrammar \citep{dong-etal-2025-xgrammar} variant enforces the same JSON schema and the label set with a grammar. Context-based prompting was run for one model per family (gemma-4-E2B-it, Qwen3.8-27B, and gpt-oss-20b), and XGrammar with reasoning off only, for gemma-4-E2B-it and Qwen3.8-27B.

\paragraph{Evaluation.} Hard and soft F1 are computed with the \texttt{metrics.py} of \citet{semin-etal-2026-strategies}, imported unmodified from their public repository at a pinned commit rather than reimplemented. UniversalNER and ESA-MT are scored per treebank and per file, and the subset scores are macro-averaged.

\paragraph{Infrastructure.} The models run with Hugging Face Transformers 5.14 \citep{wolf-etal-2020-transformers} and PyTorch 2.10 in bfloat16; gpt-oss uses its native MXFP4 quantization. Constrained and unconstrained tagging and context-based baselines were run on NVIDIA H200 GPUs (141\,GB), with several copies of a model sharing one GPU to increase throughput; XGrammar 0.2.8 ran on NVIDIA A100 GPUs (40\,GB), except for Qwen3.8-27B on UniversalNER, which ran on H200 GPUs. Including pilot runs and failed jobs, the experiments used about 4{,}300 GPU-hours between August and October 2026, 58\% of them on H200 GPUs and the rest on A100 GPUs; the constrained-tagging results reported here all come from the H200 runs.

\section{Additional Analysis}
\label{app:analysis}

\subsection{Runtime}
\label{app:runtime}

Because constrained and unconstrained tagging generate tokens at the same rate (\S\ref{subsec:cost}), their times differ only where their outputs differ in length. The largest relative gap, gemma-4-E2B-it without reasoning on LegalQAEval (15.6 against 2.7 seconds), comes from the unconstrained outputs, which stop after 32 tokens on average, and 99\% of which do not reproduce the passage.

\begin{figure*}[t]
\centering
\includegraphics[width=\textwidth]{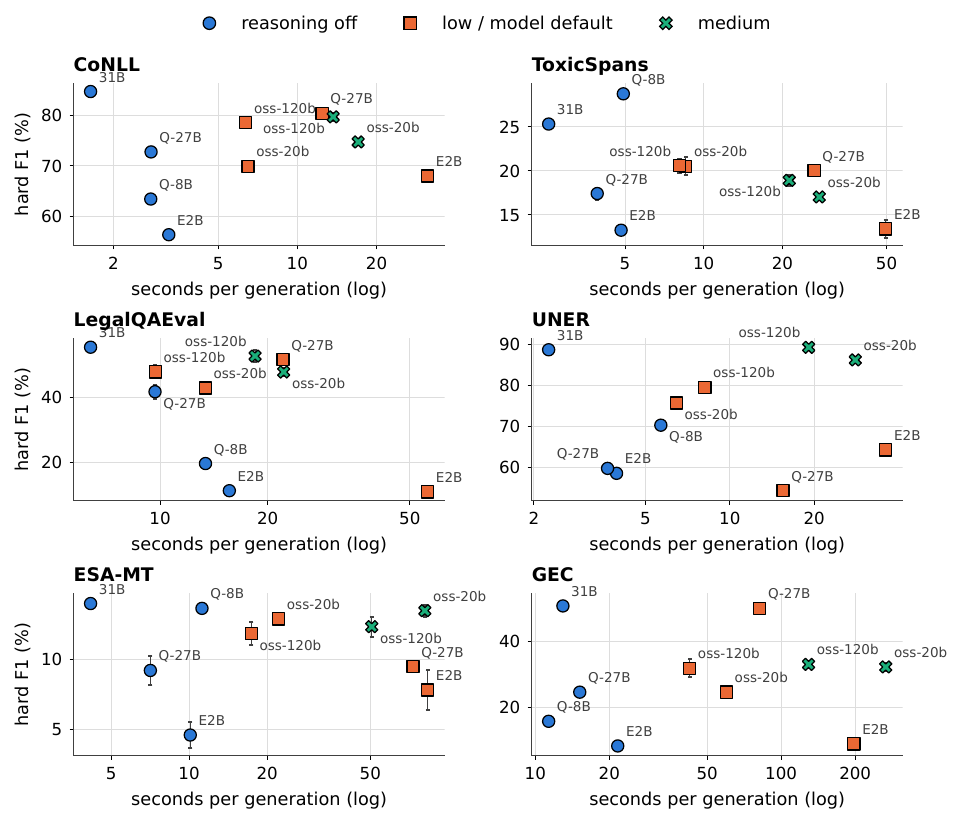}
\caption{Hard F1 against seconds per generation under constrained tagging, one point per model and reasoning setting (batch size~1, log scale); the error bars are the standard deviation over seeds.}
\label{fig:f1-time}
\end{figure*}

Across models, the time is set less by model size than by how long the model reasons (Fig.~\ref{fig:f1-time}). Without reasoning, gemma-4-31B-it is the fastest at 5.0 seconds per generation averaged over the datasets, ahead of Qwen3.8-27B (7.1), Qwen3-8B (8.2) and gemma-4-E2B-it (9.9). Throughput per process depends on how many copies of a model share a GPU: the small models ran many copies per GPU, each slower, so gemma-4-E2B-it generates 12.0 tokens per second against 25.7 for gemma-4-31B-it. Reasoning multiplies the time per generation by 7.6 for gemma-4-E2B-it (75.7 seconds with reasoning on) and by 5.5 for Qwen3.8-27B at low effort (38.6 seconds), and medium effort multiplies the time of low effort by 3.8 for gpt-oss-20b (73.7 seconds) and by 2.7 for gpt-oss-120b (42.1 seconds). MultiGEC, whose essays are the longest inputs, is the slowest dataset for every arm except Qwen3-8B, up to 266 seconds per generation for gpt-oss-20b at medium effort. With reasoning, a single generation can take from under a second to more than 2{,}000 seconds, when the reasoning runs toward the token budget.

\begin{figure*}[t]
\centering
\includegraphics[width=\textwidth]{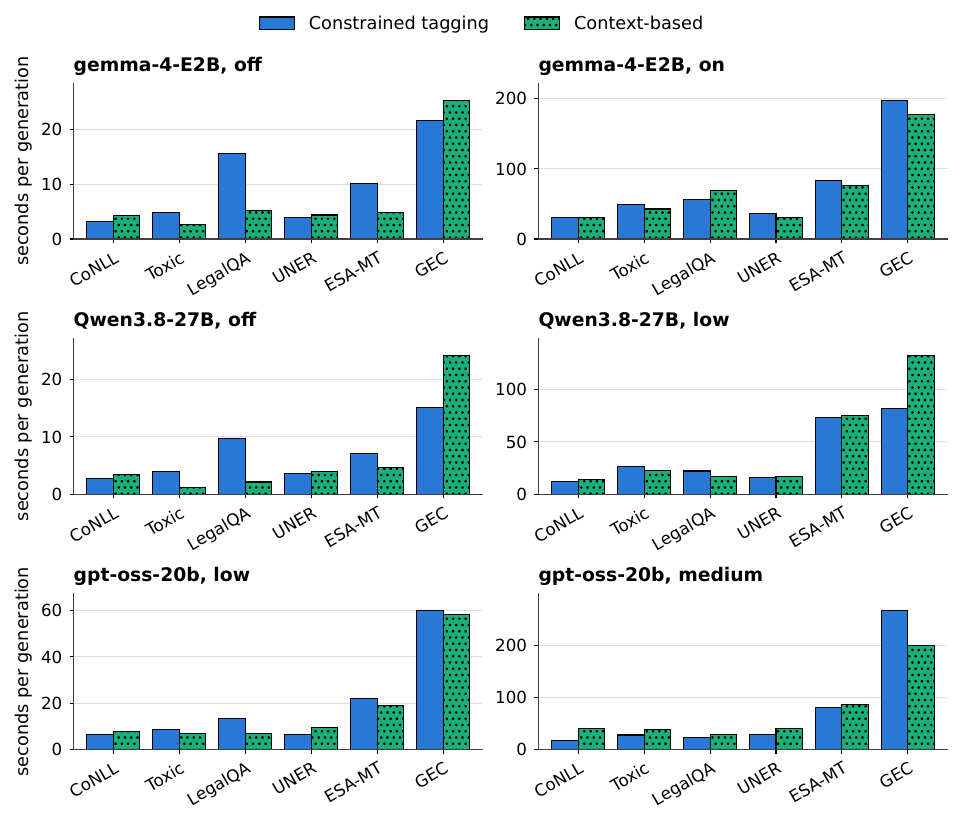}
\caption{Seconds per generation of constrained tagging and the context-based baseline on the arms run for the baseline, mean over three seeds at batch size~1, all on H200 GPUs.}
\label{fig:time-methods}
\end{figure*}

\begin{figure*}[t]
\centering
\includegraphics[width=\textwidth]{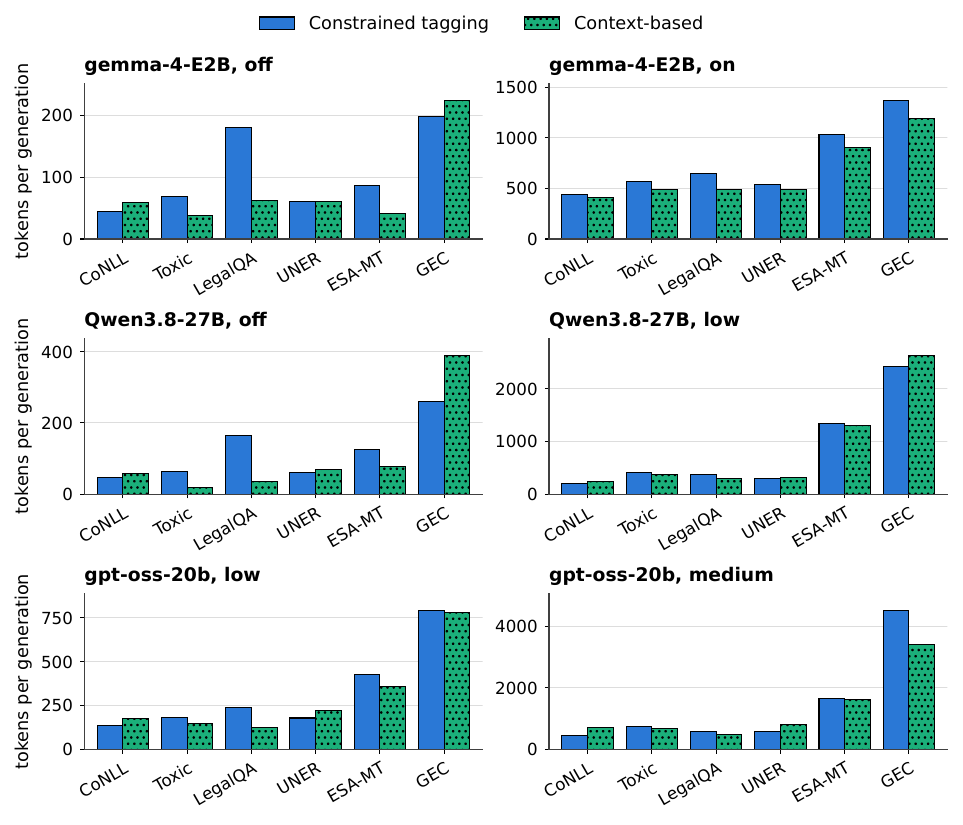}
\caption{Tokens per generation, reasoning included, of constrained tagging and the context-based baseline on the same arms as Fig.~\ref{fig:time-methods}; mean over three seeds at batch size~1. Unlike time, the token count does not depend on the hardware.}
\label{fig:cost}
\end{figure*}

Against the context-based baseline (Fig.~\ref{fig:time-methods}), constrained tagging is slower without reasoning on Toxic Spans, LegalQAEval, and ESA-MT, where it copies a long input to mark a few spans, and faster on MultiGEC, where the list of spans is longer than the essay; on the two NER datasets, the two are close. With reasoning, the reasoning dominates both, and neither method is consistently faster. Both ran on H200 GPUs, but as separate jobs with a different number of copies per GPU, so the comparison is indicative; the number of generated tokens does not depend on the hardware (Fig.~\ref{fig:cost}). More time does not buy accuracy uniformly (Fig.~\ref{fig:f1-time}): gemma-4-31B-it without reasoning has both the lowest time and the highest hard F1 averaged over the datasets.

\subsection{Reasoning}

\begin{figure*}[t]
\centering
\includegraphics[width=\textwidth]{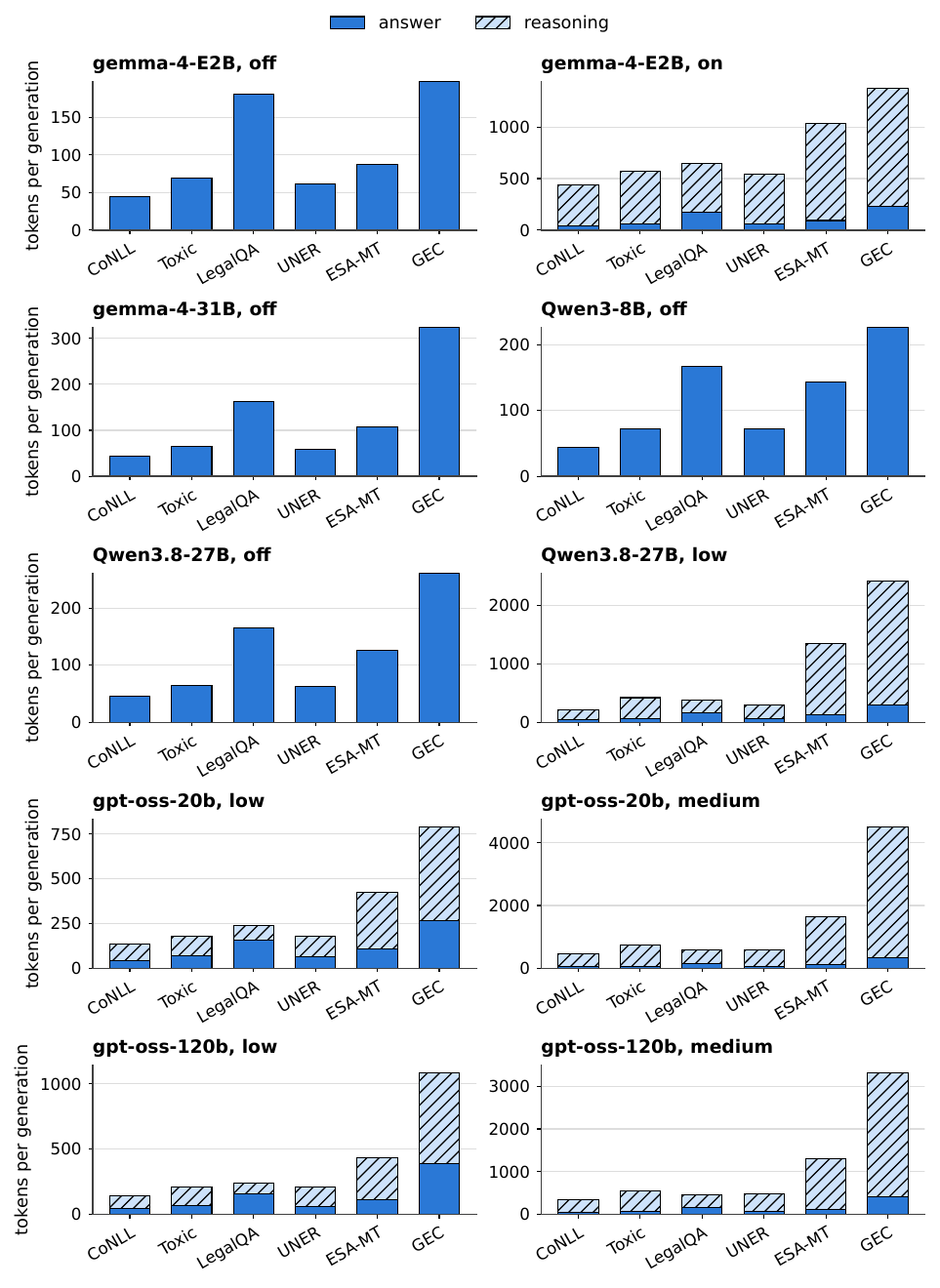}
\caption{Generated tokens per generation under constrained tagging, split into reasoning and answer, for every model and reasoning setting (batch size~1). The answer is a copy of the input plus tags and is nearly constant within a dataset; everything above it is reasoning.}
\label{fig:tokens}
\end{figure*}

Table~\ref{tab:reasoning} breaks the effect of reasoning down by dataset. Under constrained tagging, the gain is uneven: it is largest on MultiGEC for Qwen3.8-27B and on UniversalNER for gpt-oss, and it turns negative on Toxic Spans for gpt-oss and on UniversalNER for Qwen3.8-27B. Fig.~\ref{fig:tokens} shows where the generated tokens go: the answer is nearly constant within a dataset, and everything above it is reasoning.

\subsection{Failures}

Table~\ref{tab:failures} accounts for every constrained generation whose output differs from its input. None is a copying error: either the model never closed its reasoning, so there was no answer for the constraint to govern, or, at batch size~64, the token budget ran out while the answer was being copied, which leaves a correct but incomplete copy.

\section{Additional Tables}
\label{app:tables}

Table~\ref{tab:full} gives every arm of every method with its standard deviation over seeds, and the averages over the arms run with all three methods that \S\ref{subsec:main-results} quotes. Table~\ref{tab:failures} lists the failed constrained generations by cause, and Tables~\ref{tab:spans} and~\ref{tab:confusion} decompose the span and label errors discussed in \S\ref{subsec:analysis}. Table~\ref{tab:reasoning} compares the reasoning settings, and Tables~\ref{tab:batch} and~\ref{tab:batch-recall} compare batch sizes~1 and~64.

\begin{table*}[t]
\centering\footnotesize
\setlength{\tabcolsep}{3pt}
\begin{tabular}{@{}ll cccccc@{}}
\toprule
Model & Reasoning & CoNLL & Toxic Spans & LegalQAEval & UNER$^{\dagger}$ & ESA-MT$^{\dagger}$ & MultiGEC \\
\midrule
\multicolumn{8}{@{}l}{\emph{Constrained tagging (ours)}} \\
gemma-4-E2B-it & off    & 56.39\,{\scriptsize$\pm$0.64} & 13.24\,{\scriptsize$\pm$0.57} & 11.25\,{\scriptsize$\pm$0.35} & 58.58\,{\scriptsize$\pm$0.59} & 4.60\,{\scriptsize$\pm$0.95} & 8.25\,{\scriptsize$\pm$0.73} \\
               & on     & 68.00\,{\scriptsize$\pm$0.79} & 13.39\,{\scriptsize$\pm$1.03} & 10.91\,{\scriptsize$\pm$0.42} & 64.29\,{\scriptsize$\pm$0.52} & 7.80\,{\scriptsize$\pm$1.41} & 8.89\,{\scriptsize$\pm$0.67} \\
gemma-4-31B-it & off    & \textit{84.69}\,{\scriptsize$\pm$0.13} & 25.30\,{\scriptsize$\pm$0.19} & \textit{55.55}\,{\scriptsize$\pm$0.34} & 88.68\,{\scriptsize$\pm$0.14} & \textbf{13.96}\,{\scriptsize$\pm$0.16} & \textit{50.65}\,{\scriptsize$\pm$1.41} \\
Qwen3-8B       & off    & 63.44\,{\scriptsize$\pm$0.31} & 28.74\,{\scriptsize$\pm$0.33} & 19.64\,{\scriptsize$\pm$0.74} & 70.30\,{\scriptsize$\pm$0.35} & 13.62\,{\scriptsize$\pm$0.07} & 15.71\,{\scriptsize$\pm$0.35} \\
Qwen3.8-27B    & off    & 72.76\,{\scriptsize$\pm$0.09} & 17.41\,{\scriptsize$\pm$0.68} & 41.81\,{\scriptsize$\pm$2.19} & 59.78\,{\scriptsize$\pm$0.89} & 9.19\,{\scriptsize$\pm$1.04} & 24.54\,{\scriptsize$\pm$0.89} \\
               & low    & 80.39\,{\scriptsize$\pm$0.57} & 20.02\,{\scriptsize$\pm$0.51} & 51.75\,{\scriptsize$\pm$0.88} & 54.36\,{\scriptsize$\pm$0.62} & 9.46\,{\scriptsize$\pm$0.36} & 49.92\,{\scriptsize$\pm$0.18} \\
gpt-oss-20b    & low    & 69.90\,{\scriptsize$\pm$1.04} & 20.51\,{\scriptsize$\pm$1.01} & 42.94\,{\scriptsize$\pm$1.29} & 75.72\,{\scriptsize$\pm$1.36} & 12.87\,{\scriptsize$\pm$0.12} & 24.54\,{\scriptsize$\pm$0.82} \\
               & medium & 74.73\,{\scriptsize$\pm$0.49} & 17.02\,{\scriptsize$\pm$0.25} & 47.89\,{\scriptsize$\pm$1.25} & 86.20\,{\scriptsize$\pm$0.20} & 13.45\,{\scriptsize$\pm$0.41} & 32.17\,{\scriptsize$\pm$1.10} \\
gpt-oss-120b   & low    & 78.55\,{\scriptsize$\pm$0.46} & 20.57\,{\scriptsize$\pm$0.79} & 47.92\,{\scriptsize$\pm$2.00} & 79.52\,{\scriptsize$\pm$0.79} & 11.82\,{\scriptsize$\pm$0.80} & 31.76\,{\scriptsize$\pm$2.71} \\
               & medium & 79.69\,{\scriptsize$\pm$0.73} & 18.91\,{\scriptsize$\pm$0.61} & 52.81\,{\scriptsize$\pm$1.83} & \textit{89.27}\,{\scriptsize$\pm$0.30} & 12.31\,{\scriptsize$\pm$0.72} & 32.94\,{\scriptsize$\pm$1.17} \\
\midrule
\multicolumn{8}{@{}l}{\emph{Unconstrained tagging}} \\
gemma-4-E2B-it & off    & 11.31\,{\scriptsize$\pm$3.07} & 28.66\,{\scriptsize$\pm$0.53} & 2.05\,{\scriptsize$\pm$1.11} & 22.18\,{\scriptsize$\pm$0.63} & 2.00\,{\scriptsize$\pm$0.39} & 0.70\,{\scriptsize$\pm$0.27} \\
               & on     & 47.64\,{\scriptsize$\pm$0.40} & 27.62\,{\scriptsize$\pm$0.80} & 2.33\,{\scriptsize$\pm$1.32} & 69.41\,{\scriptsize$\pm$0.19} & 4.89\,{\scriptsize$\pm$1.13} & 10.32\,{\scriptsize$\pm$1.22} \\
gemma-4-31B-it & off    & 72.39\,{\scriptsize$\pm$0.18} & 24.89\,{\scriptsize$\pm$0.77} & \textbf{55.96}\,{\scriptsize$\pm$0.45} & 84.54\,{\scriptsize$\pm$0.25} & \textit{13.89}\,{\scriptsize$\pm$0.40} & 36.25\,{\scriptsize$\pm$0.86} \\
Qwen3-8B       & off    & 59.07\,{\scriptsize$\pm$1.37} & \textbf{32.62}\,{\scriptsize$\pm$0.38} & 17.93\,{\scriptsize$\pm$0.76} & 62.54\,{\scriptsize$\pm$0.13} & 13.35\,{\scriptsize$\pm$0.77} & 13.77\,{\scriptsize$\pm$0.25} \\
Qwen3.8-27B    & off    & 69.89\,{\scriptsize$\pm$0.15} & 23.31\,{\scriptsize$\pm$1.06} & 46.40\,{\scriptsize$\pm$0.76} & 60.69\,{\scriptsize$\pm$0.42} & 9.15\,{\scriptsize$\pm$0.52} & 31.14\,{\scriptsize$\pm$0.74} \\
               & low    & 83.01\,{\scriptsize$\pm$0.59} & 21.25\,{\scriptsize$\pm$0.27} & 52.94\,{\scriptsize$\pm$0.91} & 70.63\,{\scriptsize$\pm$0.90} & 9.99\,{\scriptsize$\pm$0.39} & 46.17\,{\scriptsize$\pm$1.25} \\
gpt-oss-20b    & low    & 59.79\,{\scriptsize$\pm$2.97} & 23.74\,{\scriptsize$\pm$0.96} & 40.01\,{\scriptsize$\pm$0.89} & 70.03\,{\scriptsize$\pm$0.35} & 10.16\,{\scriptsize$\pm$0.63} & 14.17\,{\scriptsize$\pm$0.96} \\
               & medium & 70.98\,{\scriptsize$\pm$0.56} & 20.34\,{\scriptsize$\pm$0.28} & 50.35\,{\scriptsize$\pm$0.96} & 83.18\,{\scriptsize$\pm$0.56} & 10.79\,{\scriptsize$\pm$0.46} & 26.61\,{\scriptsize$\pm$2.26} \\
gpt-oss-120b   & low    & 72.91\,{\scriptsize$\pm$1.00} & 21.93\,{\scriptsize$\pm$0.80} & 46.81\,{\scriptsize$\pm$1.81} & 77.88\,{\scriptsize$\pm$0.53} & 10.35\,{\scriptsize$\pm$1.51} & 17.56\,{\scriptsize$\pm$0.91} \\
               & medium & 78.69\,{\scriptsize$\pm$0.54} & 19.18\,{\scriptsize$\pm$0.44} & 54.21\,{\scriptsize$\pm$0.52} & 87.81\,{\scriptsize$\pm$0.15} & 9.88\,{\scriptsize$\pm$0.19} & 31.31\,{\scriptsize$\pm$1.28} \\
\midrule
\multicolumn{8}{@{}l}{\emph{Context-based prompting}} \\
gemma-4-E2B-it & off    & 71.47\,{\scriptsize$\pm$0.30} & \textit{32.54}\,{\scriptsize$\pm$0.92} & 38.56\,{\scriptsize$\pm$0.44} & 78.03\,{\scriptsize$\pm$0.26} & 6.46\,{\scriptsize$\pm$0.55} & 26.98\,{\scriptsize$\pm$0.52} \\
               & on     & 73.06\,{\scriptsize$\pm$0.44} & 29.53\,{\scriptsize$\pm$0.67} & 40.30\,{\scriptsize$\pm$1.52} & 83.93\,{\scriptsize$\pm$0.48} & 8.97\,{\scriptsize$\pm$1.10} & 33.93\,{\scriptsize$\pm$0.18} \\
Qwen3.8-27B    & off    & 84.12\,{\scriptsize$\pm$0.47} & 20.38\,{\scriptsize$\pm$0.97} & 17.40\,{\scriptsize$\pm$0.36} & 86.95\,{\scriptsize$\pm$0.15} & 8.77\,{\scriptsize$\pm$0.60} & 40.15\,{\scriptsize$\pm$0.55} \\
               & low    & \textbf{87.37}\,{\scriptsize$\pm$0.16} & 28.41\,{\scriptsize$\pm$0.58} & 33.46\,{\scriptsize$\pm$0.52} & \textbf{89.43}\,{\scriptsize$\pm$0.41} & 10.14\,{\scriptsize$\pm$0.57} & \textbf{51.58}\,{\scriptsize$\pm$0.22} \\
gpt-oss-20b    & low    & 73.25\,{\scriptsize$\pm$0.52} & 30.05\,{\scriptsize$\pm$0.36} & 32.15\,{\scriptsize$\pm$0.72} & 80.94\,{\scriptsize$\pm$0.23} & 11.09\,{\scriptsize$\pm$1.14} & 32.66\,{\scriptsize$\pm$0.18} \\
               & medium & 73.13\,{\scriptsize$\pm$0.29} & 26.43\,{\scriptsize$\pm$0.33} & 38.69\,{\scriptsize$\pm$0.62} & 84.46\,{\scriptsize$\pm$0.27} & 11.24\,{\scriptsize$\pm$1.25} & 40.49\,{\scriptsize$\pm$0.99} \\
\midrule
\multicolumn{8}{@{}l}{\emph{Context-based prompting with XGrammar}} \\
gemma-4-E2B-it & off    & 71.19\,{\scriptsize$\pm$1.16} & 32.39\,{\scriptsize$\pm$0.70} & 38.50\,{\scriptsize$\pm$0.57} & 78.31\,{\scriptsize$\pm$0.25} & 6.24\,{\scriptsize$\pm$0.12} & 27.33\,{\scriptsize$\pm$0.59} \\
Qwen3.8-27B    & off    & 83.96\,{\scriptsize$\pm$0.18} & 22.18\,{\scriptsize$\pm$0.35} & 17.11\,{\scriptsize$\pm$0.55} & 86.91\,{\scriptsize$\pm$0.20} & 8.09\,{\scriptsize$\pm$0.16} & 40.51\,{\scriptsize$\pm$0.88} \\
\midrule
\multicolumn{8}{@{}l}{\emph{Mean over the six arms run with all three methods}} \\
\multicolumn{2}{@{}l}{Constrained tagging (ours)} & \textit{70.36} & 16.93 & \textbf{34.42} & \textit{66.49} & \textbf{9.56} & \textit{24.72} \\
\multicolumn{2}{@{}l}{Unconstrained tagging} & 57.10 & \textit{24.15} & 32.35 & 62.69 & 7.83 & 21.52 \\
\multicolumn{2}{@{}l}{Context-based prompting} & \textbf{77.07} & \textbf{27.89} & \textit{33.43} & \textbf{83.96} & \textit{9.45} & \textbf{37.63} \\
\multicolumn{2}{@{}l}{Ours $-$ unconstrained} & $+13.26$ & $-7.22$ & $+2.08$ & $+3.80$ & $+1.73$ & $+3.20$ \\
\multicolumn{2}{@{}l}{Ours $-$ context-based} & $-6.71$ & $-10.96$ & $+1.00$ & $-17.47$ & $+0.12$ & $-12.91$ \\
\bottomrule
\end{tabular}
\caption{Hard F1 for every model, reasoning setting and method, mean $\pm$ standard deviation over three seeds, batch size~1, at the precision of the analysis (the main-text tables round to one decimal). $^{\dagger}$Macro average over treebanks or files. On CoNLL-2003 and UniversalNER the standard deviation also includes the variance of the per-seed evaluation subsample. The last block averages the six arms that were run with all three methods (gemma-4-E2B-it off and on, Qwen3.8-27B off and low, gpt-oss-20b low and medium). The best result per dataset is in bold and the second best in italics, and likewise among the three averages of the last block.}
\label{tab:full}
\end{table*}

\begin{table*}[t]
\centering\footnotesize
\begin{tabular}{@{}lrrrrrr@{}}
\toprule
 & & & \multicolumn{2}{c}{reasoning never closed} & & \\
\cmidrule(lr){4-5}
Dataset & Generations & Wrong text & token cap & model stopped & Cap in answer & Scored 1.0 \\
\midrule
CoNLL-2003   &  34{,}560 &  0 & 0 & 0 & 0 & --- \\
Toxic Spans  &  30{,}000 &  2 & 0 & 2 & 0 &   1 \\
LegalQAEval  &  36{,}180 &  3 & 2 & 1 & 0 &   1 \\
UniversalNER & 111{,}060 &  2 & 0 & 2 & 0 & --- \\
WMT24 ESA-MT &  26{,}010 &  1 & 0 & 1 & 0 &   0 \\
MultiGEC     &  15{,}120 &  2 & 2 & 0 & 0 &   0 \\
\bottomrule
\end{tabular}
\caption{Every constrained generation that did not reproduce its input, by cause, pooled over models, reasoning settings and seeds at batch size~1. The answer itself never ran out of budget. Reasoning failed to close in ten generations: four reached the cap of 18{,}000 new tokens, and in six the model ended its turn while still reasoning. \emph{Scored 1.0} counts the failures that per-example character F1 nevertheless scores as perfect because the gold is empty; it does not apply to the NER datasets. At batch size~64, 46 further generations on UniversalNER failed in the same ways, 28 of them by exhausting the budget during the answer.}
\label{tab:failures}
\end{table*}

\begin{table*}[t]
\centering\footnotesize
\begin{tabular}{@{}l cc cccccc@{}}
\toprule
& & & \multicolumn{5}{c}{\% of predicted spans} & \% of gold \\
\cmidrule(lr){4-8}\cmidrule(l){9-9}
& Spans/ex. & Length & Exact & Label & Boundary & Both & Spurious & Missed \\
\midrule
\multicolumn{9}{@{}l}{\emph{CoNLL-2003}} \\
Gold          & 1.64 & 9.5 & & & & & & \\
Ours          & 1.86 & 10.9 & 58.9 & 10.2 & 12.8 & 3.4 & 14.7 & 2.9 \\
Unconstrained & 1.24 & 9.8 & 67.7 & 10.6 & 4.5 & 2.4 & 14.8 & 35.6 \\
Context-based & 1.79 & 9.9 & 70.1 & 9.8 & 3.7 & 2.6 & 13.7 & 6.1 \\
\midrule
\multicolumn{9}{@{}l}{\emph{Toxic Spans}} \\
Gold          & 0.58 & 6.9 & & & & & & \\
Ours          & 1.23 & 22.4 & 14.8 & --- & 16.4 & --- & 68.8 & 30.9 \\
Unconstrained & 0.81 & 17.1 & 18.8 & --- & 17.8 & --- & 63.4 & 47.2 \\
Context-based & 1.07 & 12.1 & 20.7 & --- & 12.5 & --- & 66.8 & 37.4 \\
\midrule
\multicolumn{9}{@{}l}{\emph{LegalQAEval}} \\
Gold          & 0.54 & 62.8 & & & & & & \\
Ours          & 1.13 & 138.2 & 7.2 & --- & 25.9 & --- & 66.9 & 31.5 \\
Unconstrained & 0.42 & 57.1 & 27.3 & --- & 31.1 & --- & 41.6 & 55.6 \\
Context-based & 0.61 & 34.6 & 24.0 & --- & 31.4 & --- & 44.6 & 39.1 \\
\midrule
\multicolumn{9}{@{}l}{\emph{UniversalNER}} \\
Gold          & 1.92 & 8.6 & & & & & & \\
Ours          & 2.00 & 12.6 & 60.5 & 3.1 & 19.1 & 2.0 & 15.3 & 10.9 \\
Unconstrained & 1.29 & 8.9 & 76.9 & 3.4 & 9.0 & 1.6 & 9.1 & 39.3 \\
Context-based & 1.88 & 8.7 & 79.2 & 3.2 & 7.6 & 1.7 & 8.4 & 11.4 \\
\midrule
\multicolumn{9}{@{}l}{\emph{ESA-MT}} \\
Gold          & 2.21 & 13.7 & & & & & & \\
Ours          & 1.88 & 20.5 & 2.9 & 3.8 & 13.1 & 16.1 & 64.2 & 66.8 \\
Unconstrained & 0.98 & 17.1 & 4.6 & 5.8 & 14.2 & 17.2 & 58.2 & 80.6 \\
Context-based & 1.36 & 14.9 & 4.5 & 6.3 & 13.0 & 17.3 & 59.0 & 73.7 \\
\midrule
\multicolumn{9}{@{}l}{\emph{MultiGEC}} \\
Gold          & 14.09 & 4.2 & & & & & & \\
Ours          & 6.85 & 18.5 & 36.1 & 1.9 & 11.4 & 9.6 & 41.1 & 60.7 \\
Unconstrained & 3.75 & 4.4 & 41.6 & 2.0 & 9.0 & 7.8 & 39.5 & 83.7 \\
Context-based & 8.25 & 5.6 & 34.6 & 1.8 & 11.7 & 8.2 & 43.7 & 66.6 \\
\bottomrule
\end{tabular}
\caption{Where the predicted spans go wrong, on the six arms run with all three methods (batch size~1, pooled over arms and seeds). \emph{Spans/ex.}: spans per example; \emph{Length}: mean span length in characters. Every predicted span is \emph{exact} (boundaries and label right), has the right boundaries but the wrong \emph{label}, the right label but the wrong \emph{boundary}, \emph{both} wrong, or is \emph{spurious} (overlaps no gold span); \emph{missed} gold spans overlap no prediction. Toxic Spans and LegalQAEval have a single label. An unconstrained output that does not reproduce its input contributes no spans.}
\label{tab:spans}
\end{table*}

\begin{table*}[t]
\centering\footnotesize
\setlength{\tabcolsep}{3pt}
\begin{tabular}[t]{@{}lcccc@{}}
\multicolumn{5}{@{}l}{\emph{CoNLL-2003}} \\
\toprule
& \textsc{loc} & \textsc{misc} & \textsc{org} & \textsc{per} \\
\midrule
\textsc{loc} & 96.4 & 1.3 & 2.2 & 0.2 \\
\textsc{misc} & 11.5 & 79.5 & 6.7 & 2.2 \\
\textsc{org} & 31.7 & 2.9 & 64.5 & 0.9 \\
\textsc{per} & 0.7 & 1.4 & 0.6 & 97.2 \\
\bottomrule
\end{tabular}\hfill
\begin{tabular}[t]{@{}lccc@{}}
\multicolumn{4}{@{}l}{\emph{UniversalNER}} \\
\toprule
& \textsc{loc} & \textsc{org} & \textsc{per} \\
\midrule
\textsc{loc} & 97.2 & 2.2 & 0.6 \\
\textsc{org} & 10.6 & 87.5 & 1.9 \\
\textsc{per} & 1.2 & 1.2 & 97.6 \\
\bottomrule
\end{tabular}\hfill
\begin{tabular}[t]{@{}lcc@{}}
\multicolumn{3}{@{}l}{\emph{ESA-MT}} \\
\toprule
& \textsc{major} & \textsc{minor} \\
\midrule
\textsc{major} & 56.7 & 43.3 \\
\textsc{minor} & 71.8 & 28.2 \\
\bottomrule
\end{tabular}\hfill
\begin{tabular}[t]{@{}lccc@{}}
\multicolumn{4}{@{}l}{\emph{MultiGEC}} \\
\toprule
& \textsc{m} & \textsc{r} & \textsc{u} \\
\midrule
\textsc{m} & 100.0 & 0.0 & 0.0 \\
\textsc{r} & 0.0 & 93.7 & 6.3 \\
\textsc{u} & 0.0 & 20.5 & 79.5 \\
\bottomrule
\end{tabular}
\caption{Label confusion of constrained tagging on the spans whose boundaries are exactly right, pooled over all models, reasoning settings and seeds at batch size~1: rows are gold labels, columns predicted labels, in \% of the gold label. Toxic Spans and LegalQAEval have a single label.}
\label{tab:confusion}
\end{table*}

\begin{table*}[t]
\centering\footnotesize
\begin{tabular}{@{}ll cccccc@{}}
\toprule
Model & Reasoning & CoNLL & Toxic Spans & LegalQAEval & UNER$^{\dagger}$ & ESA-MT$^{\dagger}$ & MultiGEC \\
\midrule
\multicolumn{8}{@{}l}{\emph{Constrained tagging (ours)}} \\
gemma-4-E2B-it & off      & 56.4 & 13.2 & \textbf{11.2} & 58.6 & 4.6 & 8.2 \\
               & on       & \textbf{68.0} & \textbf{13.4} & 10.9 & \textbf{64.3} & \textbf{7.8} & \textbf{8.9} \\
               & $\Delta$ & +11.6 & +0.2 & $-$0.3 & +5.7 & +3.2 & +0.6 \\[2pt]
Qwen3.8-27B    & off      & 72.8 & 17.4 & 41.8 & \textbf{59.8} & 9.2 & 24.5 \\
               & low      & \textbf{80.4} & \textbf{20.0} & \textbf{51.8} & 54.4 & \textbf{9.5} & \textbf{49.9} \\
               & $\Delta$ & +7.6 & +2.6 & +9.9 & $-$5.4 & +0.3 & +25.4 \\[2pt]
gpt-oss-20b    & low      & 69.9 & \textbf{20.5} & 42.9 & 75.7 & 12.9 & 24.5 \\
               & medium   & \textbf{74.7} & 17.0 & \textbf{47.9} & \textbf{86.2} & \textbf{13.5} & \textbf{32.2} \\
               & $\Delta$ & +4.8 & $-$3.5 & +5.0 & +10.5 & +0.6 & +7.6 \\[2pt]
gpt-oss-120b   & low      & 78.6 & \textbf{20.6} & 47.9 & 79.5 & 11.8 & 31.8 \\
               & medium   & \textbf{79.7} & 18.9 & \textbf{52.8} & \textbf{89.3} & \textbf{12.3} & \textbf{32.9} \\
               & $\Delta$ & +1.1 & $-$1.7 & +4.9 & +9.8 & +0.5 & +1.2 \\
\midrule
\multicolumn{8}{@{}l}{\emph{Context-based prompting}} \\
gemma-4-E2B-it & off      & 71.5 & \textbf{32.5} & 38.6 & 78.0 & 6.5 & 27.0 \\
               & on       & \textbf{73.1} & 29.5 & \textbf{40.3} & \textbf{83.9} & \textbf{9.0} & \textbf{33.9} \\
               & $\Delta$ & +1.6 & $-$3.0 & +1.7 & +5.9 & +2.5 & +6.9 \\[2pt]
Qwen3.8-27B    & off      & 84.1 & 20.4 & 17.4 & 86.9 & 8.8 & 40.1 \\
               & low      & \textbf{87.4} & \textbf{28.4} & \textbf{33.5} & \textbf{89.4} & \textbf{10.1} & \textbf{51.6} \\
               & $\Delta$ & +3.2 & +8.0 & +16.1 & +2.5 & +1.4 & +11.4 \\[2pt]
gpt-oss-20b    & low      & \textbf{73.3} & \textbf{30.0} & 32.1 & 80.9 & 11.1 & 32.7 \\
               & medium   & 73.1 & 26.4 & \textbf{38.7} & \textbf{84.5} & \textbf{11.2} & \textbf{40.5} \\
               & $\Delta$ & $-$0.1 & $-$3.6 & +6.5 & +3.5 & +0.1 & +7.8 \\
\bottomrule
\end{tabular}
\caption{Hard F1 per dataset for every model run with two reasoning settings, and the gain from more reasoning ($\Delta$), mean over three seeds at batch size~1. gpt-oss cannot switch reasoning off, so its two settings are low and medium effort. $^{\dagger}$Macro average over treebanks or files. The context-based baseline was not run with gpt-oss-120b. For each model and method, the better reasoning setting is in bold.}
\label{tab:reasoning}
\end{table*}

\begin{table*}[t]
\centering\footnotesize
\begin{tabular}{@{}ll cccc c cccc@{}}
\toprule
& & \multicolumn{4}{c}{Constrained tagging (ours)} & & \multicolumn{4}{c}{Context-based prompting} \\
\cmidrule(lr){3-6}\cmidrule(lr){8-11}
& & \multicolumn{2}{c}{CoNLL} & \multicolumn{2}{c}{UNER$^{\dagger}$} & & \multicolumn{2}{c}{CoNLL} & \multicolumn{2}{c}{UNER$^{\dagger}$} \\
\cmidrule(lr){3-4}\cmidrule(lr){5-6}\cmidrule(lr){8-9}\cmidrule(lr){10-11}
Model & Reasoning & 1 & 64 & 1 & 64 & & 1 & 64 & 1 & 64 \\
\midrule
gemma-4-E2B-it & off    & 56.4 & 8.5 & 58.6 & 29.0 &  & \textbf{71.5} & \textbf{50.5} & \textbf{78.0} & \textbf{44.9} \\
               & on     & 68.0 & 12.2 & 64.3 & 33.7 &  & \textbf{73.1} & \textbf{58.6} & \textbf{83.9} & \textbf{59.9} \\
gemma-4-31B-it & off    & 84.7 & 86.6 & 88.7 & 86.9 &  & --- & --- & --- & --- \\
Qwen3-8B       & off    & 63.4 & 63.7 & 70.3 & 69.3 &  & --- & --- & --- & --- \\
Qwen3.8-27B    & off    & 72.8 & 69.2 & 59.8 & 63.4 &  & \textbf{84.1} & \textbf{86.6} & \textbf{86.9} & \textbf{85.9} \\
               & low    & 80.4 & 66.7 & 54.4 & 63.3 &  & \textbf{87.4} & \textbf{87.6} & \textbf{89.4} & \textbf{68.0} \\
gpt-oss-20b    & low    & 69.9 & 23.0 & 75.7 & \textbf{43.2} &  & \textbf{73.3} & \textbf{26.0} & \textbf{80.9} & 23.5 \\
               & medium & \textbf{74.7} & \textbf{64.7} & \textbf{86.2} & \textbf{53.7} &  & 73.1 & 59.9 & 84.5 & 44.5 \\
gpt-oss-120b   & low    & 78.6 & 57.6 & 79.5 & 71.8 &  & --- & --- & --- & --- \\
               & medium & 79.7 & 78.5 & 89.3 & 80.3 &  & --- & --- & --- & --- \\
\bottomrule
\end{tabular}
\caption{Hard F1 at batch size~1 and~64 on the two NER datasets, mean over three seeds. $^{\dagger}$Macro average over the 18 treebanks. Marks a model for which the context-based baseline was not run. Copying 64 sentences also strains the token budget: with reasoning, 13.7\% of the batched UNER generations of Qwen3.8-27B (low) and 9.8\% of gpt-oss-20b (medium) did not complete (Table~\ref{tab:failures}). For each model, reasoning setting, dataset and batch size, the better of the two methods is in bold.}
\label{tab:batch}
\end{table*}

\begin{table*}[t]
\centering\footnotesize
\setlength{\tabcolsep}{3.5pt}
\begin{tabular}{@{}ll cccccc c cccccc@{}}
\toprule
& & \multicolumn{6}{c}{Constrained tagging (ours)} & & \multicolumn{6}{c}{Context-based prompting} \\
\cmidrule(lr){3-8}\cmidrule(lr){10-15}
& & \multicolumn{3}{c}{CoNLL} & \multicolumn{3}{c}{UNER$^{\dagger}$} & & \multicolumn{3}{c}{CoNLL} & \multicolumn{3}{c}{UNER$^{\dagger}$} \\
\cmidrule(lr){3-5}\cmidrule(lr){6-8}\cmidrule(lr){10-12}\cmidrule(lr){13-15}
Model & Reasoning & R$_1$ & R$_{64}$ & P$_{64}$ & R$_1$ & R$_{64}$ & P$_{64}$ & & R$_1$ & R$_{64}$ & P$_{64}$ & R$_1$ & R$_{64}$ & P$_{64}$ \\
\midrule
gemma-4-E2B-it & off    & 69.9 & 20.2 & 5.3 & \textbf{77.0} & \textbf{46.3} & 26.3 &  & \textbf{73.5} & \textbf{36.8} & \textbf{81.0} & 74.3 & 32.6 & \textbf{81.7} \\
               & on     & \textbf{80.1} & 28.9 & 7.7 & 83.7 & \textbf{51.2} & 30.8 &  & 79.3 & \textbf{45.3} & \textbf{82.9} & \textbf{84.6} & 46.9 & \textbf{88.6} \\
gemma-4-31B-it & off    & 88.5 & 88.1 & 85.2 & 89.6 & 85.2 & 89.8 &  & --- & --- & --- & --- & --- & --- \\
Qwen3-8B       & off    & 64.8 & 59.0 & 69.6 & 78.3 & 65.6 & 77.6 &  & --- & --- & --- & --- & --- & --- \\
Qwen3.8-27B    & off    & 84.8 & 63.0 & 77.1 & 79.1 & 55.6 & 83.5 &  & \textbf{86.5} & \textbf{85.2} & \textbf{88.1} & \textbf{87.5} & \textbf{83.9} & \textbf{88.5} \\
               & low    & 84.1 & 57.7 & 80.2 & 80.5 & 55.0 & 84.3 &  & \textbf{89.4} & \textbf{85.3} & \textbf{90.1} & \textbf{89.1} & \textbf{60.1} & \textbf{90.0} \\
gpt-oss-20b    & low    & \textbf{82.0} & \textbf{21.5} & 25.7 & \textbf{88.3} & \textbf{46.9} & 49.5 &  & 79.9 & 15.6 & \textbf{79.1} & 81.0 & 17.6 & \textbf{59.3} \\
               & medium & \textbf{83.5} & \textbf{65.6} & 63.9 & \textbf{91.1} & \textbf{57.9} & 58.4 &  & 83.1 & 48.0 & \textbf{79.8} & 86.3 & 34.7 & \textbf{78.8} \\
gpt-oss-120b   & low    & 84.2 & 48.8 & 70.8 & 89.3 & 68.5 & 79.2 &  & --- & --- & --- & --- & --- & --- \\
               & medium & 84.4 & 75.5 & 81.8 & 90.8 & 78.1 & 84.0 &  & --- & --- & --- & --- & --- & --- \\
\bottomrule
\end{tabular}
\caption{Recall at batch size~1 and~64 (R$_1$, R$_{64}$) and precision at batch size~64 (P$_{64}$) behind Table~\ref{tab:batch}: the hard character-overlap metric, mean over three seeds. $^{\dagger}$Macro average over the 18 treebanks. --- marks a model the context-based baseline was not run with. For each model, reasoning setting, dataset and column, the better of the two methods is in bold.}
\label{tab:batch-recall}
\end{table*}

\section{System Prompts}
\label{app:prompts}

Every prompt follows one structure: a task description, the label set with a definition of each label, the output format, a list of rules, and one or two worked examples. Listing~\ref{lst:tagging} gives the complete tagging prompt for CoNLL-2003, used for both constrained and unconstrained tagging, and Table~\ref{tab:prompts} lists what changes for the other datasets. Each context-based prompt keeps the task description, labels and examples of the tagging prompt for the same dataset, with the examples written as JSON, and replaces its output format and rules with those of Listing~\ref{lst:context}; the context is four to eight words long, three to eight on Toxic Spans. All twelve prompts are released verbatim with our code. Placeholders in braces are filled per example.

\lstset{basicstyle=\ttfamily\scriptsize, breaklines=true, columns=fullflexible, keepspaces=true, frame=single, framesep=2pt, xleftmargin=3pt, xrightmargin=3pt, captionpos=t, escapeinside={(*@}{@*)}}

\begin{lstlisting}[caption={Tagging prompt for CoNLL-2003.},label={lst:tagging}]
You are an expert at named entity recognition. Given an input text, identify all named entities and return the SAME text with inline entity markup.
Do not extract nested entities, only the outermost ones.

IMPORTANT: If an entity appears multiple times, tag each occurrence.

The possible labels for named entities are:
PER: names of people (different languages, nicknames, usernames, fictional characters, titles with names, etc.)
LOC: names of cities, countries, landmarks, geographical features, addresses, etc.
ORG: names of companies, institutions, agencies, sport teams, etc.
MISC: everything else that can be considered a named entity (events, works of art, nationalities, religions, languages, etc.)

The order of labeling is PER, LOC, ORG, MISC.

Output format:
Return ONLY the input text with each entity wrapped exactly like this:
<SPAN><LABEL>LABEL</LABEL>ENTITY_TEXT</SPAN>

Rules:
- Do not add, remove, or reorder any characters from the original input text, except for inserting the tags.
- ENTITY_TEXT must exactly match the original substring from the input text.
- Do not output explanations, or any additional text!!
- Do not tag anything that is not one of the labels above.
- Do not create overlapping spans; when ambiguous, choose the outermost entity.

Examples:
Input text:
Barack Obama was born in Hawaii. Barack was american.
Output text:
<SPAN><LABEL>PER</LABEL>Barack Obama</SPAN> was born in <SPAN><LABEL>LOC</LABEL>Hawaii</SPAN>. <SPAN><LABEL>PER</LABEL>Barack</SPAN> was american.

Input text:
He ended the World Cup on the wrong note , Coste said .
Output text:
He ended the <SPAN><LABEL>MISC</LABEL>World Cup</SPAN> on the wrong note , <SPAN><LABEL>PER</LABEL>Coste</SPAN> said .
\end{lstlisting}

\begin{lstlisting}[caption={Output format and rules of the context-based prompt for CoNLL-2003.},label={lst:context}]
Output format:
Return ONLY a JSON array of objects, each exactly like this:
[
    { "entity": "ENTITY_TEXT", "label": "LABEL", "context": "SURROUNDING_CONTEXT" }
]

Rules:
- "entity" must exactly match the original substring from the input text.
- "label" must be one of the labels above.
- "context" must be a short snippet (4-8 words) copied exactly from the input text that contains the entity and a few neighbouring words.
- The entity must be contained in the context snippet.
- If the entity is at the beginning or end of the text, use only the available neighbouring words.
- Do not output explanations, or any additional text!!
- Do not tag anything that is not one of the labels above.
- Do not create overlapping spans; when ambiguous, choose the outermost entity.
- If there are no named entities, output an empty JSON array [].
\end{lstlisting}

\begin{table*}[t]
\centering\footnotesize
\begin{tabular}{@{}l p{4.1cm} p{3.5cm} p{4.6cm} c@{}}
\toprule
Dataset & Task description & Labels & Additional rules & Examples \\
\midrule
CoNLL-2003 & named entity recognition & \textsc{per}, \textsc{loc}, \textsc{org}, \textsc{misc} & outermost entities only; tag every occurrence & 2 \\
UniversalNER & named entity recognition in any language and script, without translating the input & \textsc{per}, \textsc{org}, \textsc{loc}; no \textsc{misc} & as for CoNLL-2003 & 2 \\
Toxic Spans & toxic spans in a social-media post & \textsc{toxic}: insults, slurs, threats, hate speech, identity attacks and profanity aimed at a person & return the post unchanged if nothing is toxic & 2 \\
LegalQAEval & the span of a legal passage that answers \texttt{\{question\}} & \textsc{answer} & return the passage unchanged if it has no answer; of identical answers, tag only the first & 1 \\
ESA-MT & errors in a translation, given the \texttt{\{source\_text\}} in \texttt{\{source\_language\}} as context that is never copied & \textsc{major}, \textsc{minor} (error severity) & return the translation unchanged if it has no errors & 1 \\
MultiGEC & grammatical errors in learner English & \textsc{r} (replace), \textsc{u} (unnecessary), \textsc{m} (missing) & an \textsc{m} tag is empty and marks where a word is missing; never write the correction; no two \textsc{m} tags at one position & 1 \\
\bottomrule
\end{tabular}
\caption{What the system prompt of each dataset changes relative to Listing~\ref{lst:tagging}. The output format and the general rules, such as reproducing the input exactly and not creating overlapping spans, are shared by all datasets.}
\label{tab:prompts}
\end{table*}

\section{Processor Details}
\label{app:processor}

Two implementation details complete the processor of \S\ref{sec:method-processor}. The end-of-sequence set is the union of the tokenizer's and the model's end tokens and the family's end-of-turn tokens, because admitting one while generation stops only on another runs every example to the token budget. If the admissible set is ever empty, the processor admits end-of-sequence rather than sampling from a row of $-\infty$.

\begin{figure*}[t]
\centering
\includegraphics[width=0.85\textwidth]{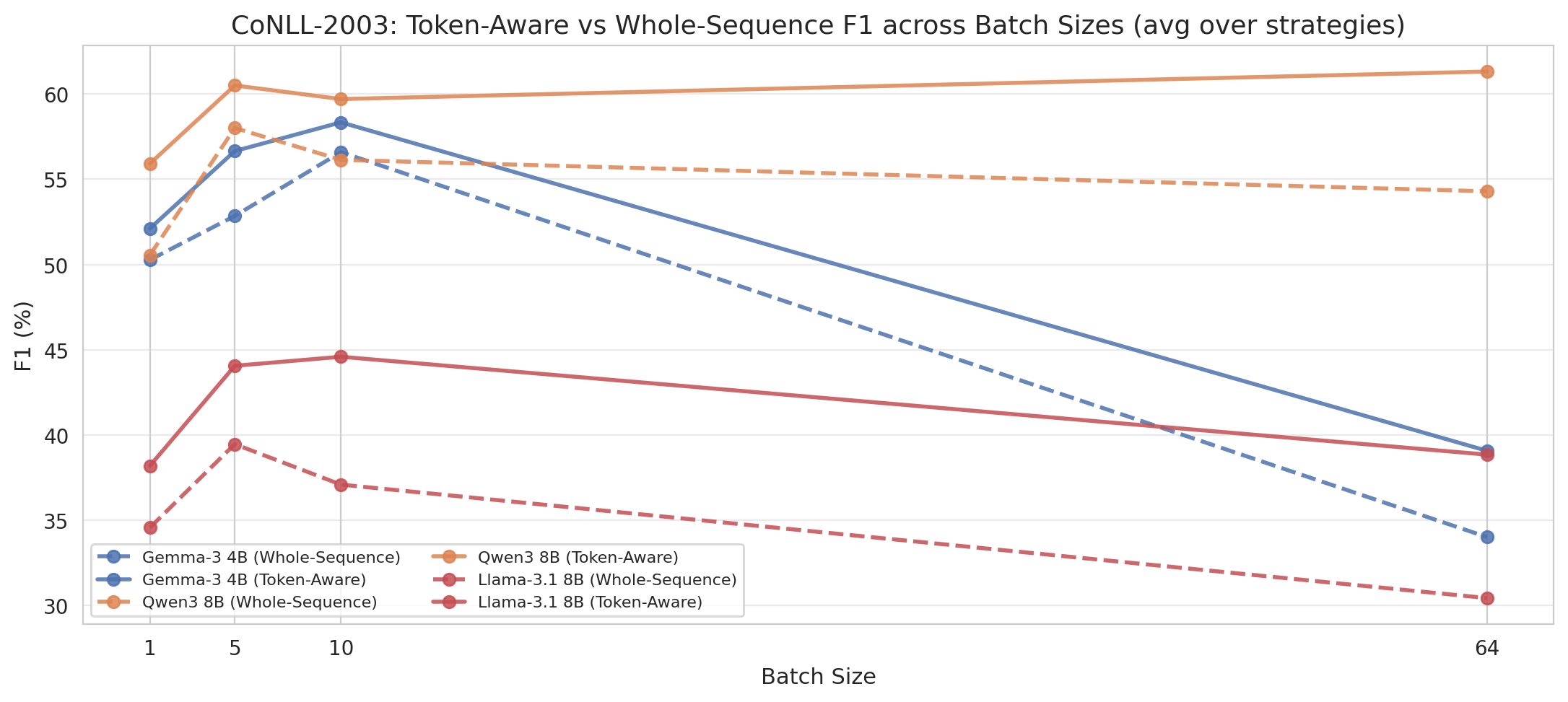}
\caption{\ta{} against \ws{} on CoNLL-2003, from earlier experiments with three smaller instruction-tuned models, averaged over greedy and sampling decoding.}
\label{fig:processor-comparison}
\end{figure*}

We implemented two variants of the processor, which differ only in how the copy cursor moves. \ta{}, used in all experiments, tracks the current input token and the byte offset inside it and searches the trie only with the rest of that token, so a copied token never crosses from one input token into the next. \ws{} keeps a single byte pointer into $x$ and searches the trie with all of the remaining input, so a copied token may cover the end of one input token together with the start of the next. Both reach every boundary, including those inside a token. In earlier experiments with three smaller instruction-tuned models, \ta{} was ahead on CoNLL-2003 for every model at every batch size, and the gap widened with the batch size; on Toxic Spans and LegalQAEval the two were within the seed spread of each other (Fig.~\ref{fig:processor-comparison}). We therefore use \ta{} throughout.

\section{Algorithms}
\label{app:algorithms}

Pseudocode for building the byte-level token trie and searching it (\S\ref{sec:method-trie}), taken from the implementation; each caption names the function it describes.

\begin{algorithm*}[t]
\footnotesize
\caption{Byte-level token trie construction (\texttt{build\_toktrie\_from\_tokenizer})}
\label{alg:trie-build}
\KwIn{tokenizer with vocabulary $\mathcal{V}$ and decoder $D$}
\KwOut{trie $\mathcal{T}$, whose nodes map a byte to a child node and store the ids of the tokens ending there; map \texttt{token\_bytes}: token id $\to$ bytes}
\tcp{Decide the vocabulary family ONCE, from the decoder.}
$\textit{byte\_level} \leftarrow$ ($D$ is, or contains, a \texttt{ByteLevel} decoder)\;
$\beta \leftarrow$ inverse of \texttt{bytes\_to\_unicode()}\tcp*{printable character $\to$ the raw byte it stands for}
$\mathcal{T} \leftarrow$ a trie holding only its root node\;
\ForEach{token id $u \in \mathcal{V}$}{
	$s \leftarrow \texttt{convert\_ids\_to\_tokens}$\;
	\uIf{\textit{byte\_level}}{
		$b \leftarrow (\beta[c])_{c \in s}$, or $\texttt{UTF8}(s)$ if some $c \notin \mathrm{dom}(\beta)$\tcp*{added special tokens are plain strings}
	}
	\uElseIf{$s$ has the form \texttt{<0x}$NN$\texttt{>}}{
		$b \leftarrow$ the single byte $NN$\tcp*{SentencePiece byte fallback}
	}
	\Else{
		$b \leftarrow \texttt{UTF8}(s$ with the word-boundary marker U+2581 replaced by a space$)$\;
	}
	$\texttt{token\_bytes}[u] \leftarrow b$\;
	$i \leftarrow$ root\;
	\For{$t \leftarrow 1$ \KwTo $|b|$}{
		\lIf{$i$ has no child for byte $b_t$}{create a node $j$ and add the edge $i \xrightarrow{b_t} j$}
		$i \leftarrow \texttt{child}(i, b_t)$\;
	}
	append $u$ to $\texttt{token\_ids}(i)$\;
}
\Return $\mathcal{T}$, \texttt{token\_bytes}\;
\end{algorithm*}

\begin{algorithm}[t]
\footnotesize
\caption{Trie prefix search (\texttt{prefix\_search})}
\label{alg:prefix}
\KwIn{remaining input bytes $r_1 \dots r_m$, trie $\mathcal{T}$}
\KwOut{set $A$ of the ids of every token whose bytes are a prefix of $r_1 \dots r_m$}
$i \leftarrow$ root; $A \leftarrow \emptyset$\;
\For{$t \leftarrow 1$ \KwTo $m$}{
	\lIf{$i$ has no child for byte $r_t$}{\textbf{break}}
	$i \leftarrow \texttt{child}(i, r_t)$\;
	$A \leftarrow A \cup \texttt{token\_ids}(i)$\;
}
\Return $A$\;
\end{algorithm}

\end{document}